\magnification1200


\vskip 2cm
\centerline {\bf $K_{27}$ as a symmetry of closed bosonic strings and branes}
\vskip 0.5cm

\centerline{Keith Glennon}
\vskip 0.5cm
\centerline{{\it Okinawa Institute for Science and Technology,}}
\centerline{{\it 1919-1 Tancha, Onna-son, Okinawa 904-0495, Japan}}
\centerline{}
\centerline{and}
\centerline{}
\centerline{Peter West}
\vskip 0.5cm
\centerline{{\it Mathematical Institute, University of Oxford,}}
\centerline{{\it Woodstock Road, Oxford, OX2 6GG, UK}}
\centerline{}
\centerline{{\it Department of Mathematics, King's College, London}}
\centerline{{\it The Strand, London WC2R 2LS, UK}}
\vskip 1cm
\centerline{keith.glennon@oist.jp, peter.west540@gmail.com}
\vskip 2cm
\leftline{\sl Abstract}  
We show that the dynamics encoded in the non-linear realisation of the semi-direct product of the very extended algebra $K_{27}$ with its vector representation contains the low energy effective action of the closed bosonic string.

\par 
\vskip2cm
\noindent

\vskip .5cm

\vfill
\eject

\medskip
{{\bf 1. introduction}}
\medskip
The low energy effective actions of the type IIA and IIB superstrings are  one of the properties of these superstrings that we know for sure, they are  the IIA or IIB supergravity theories respectively [1],[2]. The eleven dimensional supergravity theory [3] has been proposed to be the low energy effective action for a theory called M theory [4,5] but little is known about this theory other than its connections with supergravity and superstring  theories. These three supergravity theories are invariant under maximal  supersymmetry algebras which uniquely determine them. As a result we can be sure that they really do encode all the low energy effects of these string theories,  at least from the point particle perspective. 
\par
It has been shown these  three supergravity theories arises from a single theory called E theory [6,7], for a review see [8]. This theory is the non-linear realisation of the semi-direct algebra of $E_{11}$ with its vector representation [7] and it leads uniquely to the IIA and IIB   supergravity theories  in ten dimensions and the eleven dimensional supergravity theory [9,10,11]. These theories live on a very large spacetime but when one restricts this  to be usual spacetime the supergravity theories appear.   This restriction is required if one only takes account of point particles and not extended objects such as the branes [12]. By constructing the corresponding irreducible representation it was shown that the non-linear realisation  contained the same degrees of freedom as the supergravity theories [13]. 
\par
The open and closed bosonic strings exist in twenty six dimensions. They are consistent theories and they are much simpler than than the superstrings in ten dimensions. Their effects at low energy are known from string perturbation theory [14,15]. It was proposed in the first $E_{11}$ paper [6] that the low energy effective action of the closed bosonic string possessed a very large symmetry. More precisely it was the theory that arose from the non-linear realisation of the semi-direct algebra $K_{27}$  with its vector representation, denoted $K_{27}\otimes_s l_1$. The Kac-Moody algebra $K_{27}$, like $E_{11}$, is a very extended algebra,  more precisely it is $D_{24}^{+++}$. The  initial connection between the closed bosonic string and $K_{27}$ was made in [6] and a more extensive listing of the generators of $K_{27}$ at low levels was given in [16].  
\par
In this paper we compute the $K_{27}$ algebra and its  vector representation  at low levels and construct the non-linear realisation of $K_{27}\otimes_s l_1$.
At low levels this theory  contains the field $h_a{}^b$, a spin zero field $\phi$ a two form field $A_{a_1a_2}$ which we can identify with the graviton, the dilaton and the Kalb-Ramond two form respectively . At higher levels it contains their duals, namely  
$A_{a_1\ldots a_{23},b}$,  $A_{a_1\ldots a_{24}}$ and $A_{a_1\ldots a_{22}}$ respectively as well as higher dual fields at higher levels. 
\par
To find the  dynamics of the non-linear realisation  we construct  a set of expressions  that transform into each other under the symmetries of the non-linear realisation. These expressions can then be consistently set to zero to give the field equations. We find these unique  equations of motion for the graviton, the dilaton and the two form which are summarised in equations (5.3.1-3).  We also find the duality relations between the graviton, the dilaton  and the two form  field and their duals. These duality relations  are summarised in equations (4.0.1-3). Restricting the dependence of the fields to the usual  twenty six dimensional spacetime we find precise agreement with the known effective action for the closed bosonic string. If we were to construct the corresponding  irreducible representation we can expect to find that the degrees of freedom in the  theory are just the massless particles of the bosonic string.


\medskip
{{\bf 2. The Kac-Moody algebra $K_{27}$ and its vector representation}}
\medskip
We now establish the basic properties of the Kac-Moody algebra $K_{27} = D_{24}^{+++}$ and its $l_1$ representation at low levels. 
We follow the method used for $E_{11}$, see [17] for a review. The Dynkin diagram of $K_{27}$ is given by
$$
\matrix{
& & &  & & & \oplus & 26 & & & & & & & \oplus & 27 & & \cr
& & &  & & & | & & & & & & & & | & & & \cr
\bullet & - & \bullet & - & \bullet & - & \bullet & - & \bullet & - & \cdots & - & \bullet & - & \bullet & - & \bullet \cr
1 & & 2 & & 3 & & 4 & & 5 & & & & 23 & & 24 & & 25 \cr
} \eqno(2.1) 
$$ 
As with any Kac-Moody algebra that is not of finite or affine type, the full listing of the generators of $K_{27}$ is not known. If we delete nodes 26 and 27 we are left with the  sub-algebra $A_{25}$ and we can investigate the  $K_{27}$ algebra by decomposing it into this sub-algebra.  This is indicated in the diagram of equation (2.1) by the $\oplus$ signs used for nodes 26 and 27. The resulting generators can be classified in terms of two integers $l_{26}$ and $l_{27}$, associated to nodes 26 and 27, which we write as  $\vec{l} = (l_{26},l_{27})$. We will also define the combined  level of a generator by  $l = l_{26} + l_{27}$. The number of up minus down indices on a generator is equal to $22 l_{26} + 2 l_{27}$, and we can use this to determine the level of a generator by inspection. The non-negative level generators  up to level three are as follows
$$
K^a{}_b  \ \ (0,0) \ \ , \ \ R \ \ (0,0) \ \ ; \ \ R^{a_1 a_2} \ \ (0,1) \ \ , \ \ R^{a_1 ... a_{22}} \ \ (1,0) \ \ ; $$
$$
R^{a_1 ... a_{24}} \ \ (1,1) \ \ , \ \ R^{a_1 .. a_{23},b} \ \ (1,1) \ \ , \ \ R^{a_1 .. a_{25},b_1 .. b_{19}} \ \ (2,0)  \ \ ; $$
$$
R^{a_1 .. a_{25},b} \ \ (1,2) \ \ , \ \ R^{a_1 .. a_{24},b_1 b_2} \ \ (1,2) \ \ , \ \ R^{a_1 .. a_{23},b_1 b_2 b_3} \ \ (1,2) \ \ , $$
$$
R_{\{1\}}^{a_1 .. a_{26},b_1 .. b_{20}} \ \ (2,1) \ \ , \ \ R_{\{2\}}^{a_1 .. a_{26},b_1 .. b_{20}} \ \ (2,1) \ \ , \ \ R^{a_1 .. a_{26},b_1 .. b_{19},c} \ \ (2,1) \ \ , \ \  \eqno(2.2)
 $$
$$
R^{a_1 .. a_{26},b_1 .. b_{18},c_1 c_2} \ \ (2,1) \ \ , \ \ R^{a_1 .. a_{25},b_1 .. b_{21}} \ \ (2,1) \  \ , \ \ R^{a_1 .. a_{25},b_1 .. b_{20},c} \ \ (2,1) \ \ , $$
$$
R^{a_1 .. a_{24},b_1 .. b_{22}} \ \ (2,1) \ \ , \ \ R^{a_1 .. a_{26},b_1 .. b_{26},c_1 .. c_{14}} \ \ (3,0) \ \ , \ \ R^{a_1 .. a_{26},b_1 .. b_{25},c_1 .. c_{15}} \  \ (3,0) \  , $$
$$
R^{a_1 .. a_{26},b_1 .. b_{24},c_1 .. c_{16}} \  \ (3,0) \ \ ,  \ \ R^{a_1 .. a_{25},b_1 .. b_{24},c_1 .. c_{17}} \  \ (3,0) \ , \ \ R^{a_1 .. a_{26},b_1 .. b_{22},c_1 .. c_{18}} \ \  \ (3,0) \ \ ; \ldots $$
\par 
We have separated the generators of different levels by a semi-colon, and listed the corresponding level vector $\vec{l}$ beside each generator for convenience. The dots at the end of equation (2.2) represents the existence of generators at levels higher than three. The  indices are taken to be anti-symmetric in each block of indices, where blocks of indices are separated by commas. We have used subscripts in curly brackets to distinguish generators  when the multiplicity of the corresponding root is higher than one.
\par 
This list essentially agrees with the listing of $K_{27}$ to level three in Table 33 of the PhD thesis [16]. While the low level generators of  Kac-Moody algebra can be found by an analytic calculation, the very useful the program SimpLie [18] is often used. However, this can only be used to find the generators from levels $(0,0)$ up to the $(1,1)$ inclusive as the decomposition of the algebra $K_{27}$ involves representations that have a high dimension. To find the remaining generators, we have studied $K_n = D_{n-3}^{+++}$ for certain cases $n \leq 27$, and assumed the results follow the same pattern for $n=27$. This method did not however give the multiplicity of the generator  $R^{a_1 .. a_{26},b_1 .. b_{18},c_1 c_2} \ \ (2,1)$. We recognise among the generators those with  blocks of twenty four indices which are  expected from duality considerations. We will not need the higher level generators in this paper. We will  compute the $K_{27}$ algebra up to level two and the  consistency of this algebra ensures that we do indeed  have the generators listed above. 
\par 
The  generators satisfy irreducibility relations which for the positive generators  are given by
$$
R^{[a_1 .. a_{23},b]} = 0 \ ; \ R^{[a_1 .. a_{25},b_1] b_2 \ldots b_{19}} = 0 , \ldots \ \ . 
\eqno(2.3)$$
The negative root generators can be read off from equation (2.2) and those up to level two are  given by 
$$
R_{a_1 a_2} \ , \ R_{a_1 .. a_{22}} \ ; \ R_{a_1 .. a_{24}} \ , \ R_{a_1 .. a_{23},b} \ , \ R_{a_1 .. a_{25},b_1 .. b_{19}} \ , \  \ . \eqno(2.4)$$
These obey similar  irreducibility relations. 
\par 
In this paper we will construct the  $K_{27}$ algebra up and including  to level two in appendix A but , for convenience,  we give here  the commutators involving   generators with  levels 1, 0, -1.  The algebra of the non-negative generators is given by
$$
[K^a{}_b,R] = 0 \ , \ [K^a{}_b,R^{c_1 c_2}] = 2 \delta^{[c_1}{}_b R^{|a|c_2]} \ , \ [K^a{}_b,R^{c_1 .. c_{22}}] = 22 \delta^{[c_1}{}_b R^{|a|c_2 .. c_{22}]} \ , $$
$$
 [K^a{}_b,R^{c_1 .. c_{24}}] = 24 \delta^{[c_1}{}_b R^{|a|c_2 .. c_{24}]} \ ,  $$
$$
[R,R] = 0 \ \ , \ \ [R,R^{a_1 a_2}] = R^{a_1 a_2}  \ , \ [R,R^{a_1 ... a_{22}}] = - R^{a_1 ... a_{22}}  \ , \eqno(2.5) $$
$$
[R^{a_1 a_2},R^{b_1 b_2}] = 0 \ , \ [R^{a_1 a_2},R^{b_1 .. b_{22}}] = R^{a_1 a_2 b_1 .. b_{22}} + R^{b_1 ... b_{22}[a_1 , a_2]} \ , \  $$
$$
[R^{a_1 .. a_{22}},R^{b_1 .. b_{22}}] = R^{a_1 .. a_{22} [b_1 b_2 b_3,b_4 .. b_{22}]} \ .$$
\par 
The algebra among the negative generators is given by
$$
[K^a{}_b,R_{c_1 c_2}] = - 2 \delta^a{}_{[c_1} R_{|b|c_2]} \ , \ [K^a{}_b,R_{c_1 .. c_{22}}] = - 22 \delta^a{}_{[c_1} R_{|b|c_2 .. c_{22}]} \ , $$
$$
[R,R_{a_1 a_2}] = - R_{a_1 a_2} \ , \ [R,R_{a_1 ... a_{22}}] = + R_{a_1 ... a_{22}} \ , $$
$$
[R_{a_1 a_2},R_{b_1 b_2}] = 0 \ , \ [R_{a_1 a_2},R_{b_1 .. b_{22}}] = R_{a_1 a_2 b_1 .. b_{22}} + R_{b_1 ... b_{22}[a_1 , a_2] } \ , \eqno(2.6) $$
$$
[R_{a_1 .. a_{22}},R_{b_1 .. b_{22}}] = R_{a_1 .. a_{22} [b_1 b_2 b_3,b_4 .. b_{22}]}  \ .$$
The commutators involving positive and negative generators  is given by 
$$
[R^{a_1 a_2},R_{b_1 b_2}] = 4 \delta^{[a_1}{}_{[b_1} K^{a_2]}{}_{b_2]} - {1 \over 6} \delta^{a_1 a_2}_{b_1 b_2} D + {1 \over 3} \delta^{a_1 a_2}_{b_1 b_2} R  \ \ ,$$
$$
[R^{a_1 a_2},R_{b_1 ... b_{22}}] = [R_{a_1 a_2},R^{b_1 ... b_{22}}] = 0 \ \ , \eqno(2.7)
$$ 
$$
[R^{a_1 .. a_{22}},R_{b_1 .. b_{22}}] = (22)^2 \cdot 21! \delta^{[a_1 .. a_{21}}_{[b_1 .. b_{21}} K^{a_{22}]}{}_{b_{22}]} - {11 \over 12} 22! \delta^{a_1 .. a_{22}}_{b_1 .. b_{22}} D - {22! \over 6} \delta^{a_1 .. a_{22}}_{b_1 .. b_{22}} R  \ \ .$$
In appendix A we will explain how this algebra is constructed and in particular what are the Serre generators. 
\par 
We now define the Cartan involution on $K_{27}$. The Cartan involution is an involution on $K_{27}$ satisfying $I_c(AB) = I_c(A) I_c(B)$, on any two generators $A$ and $B$. It is usually  defined by its  action  on the Chevalley generators by taking $I_c(E_A) = - F_A$, $I_c(F_A) = - E_A$, $I_c(H_A) = - H_A$.  However, to take account of the fact that we are going to construct theories in Minkowski spacetime  rather than Euclidean space rather than we will use a slightly modified Cartan involution which involves   the Minkowski metric, $\eta_{ab} = (-1,1,...,1)$, to raise and lower indices. 
\par 
The action of this Cartan involution on the generators of equation (2.2) is given by 
$$
I_c(K^a{}_b) = - \eta^{ad} \eta_{bc} K^c{}_d \ , \ I_c(R) = - R \ ; \ I_c(R^{a_1 a_2}) = - \eta^{a_1 b_1} \eta^{a_2 b_2} R_{b_1 b_2} \ , $$
$$
I_c(R^{a_1 .. a_{22}}) = - \eta^{a_1 b_1} .. \eta^{a_{22} b_{22}} R_{b_1 .. b_{22}} \ ; \ I_c(R^{a_1 .. a_{24}}) = + \eta^{a_1 b_1} .. \eta^{a_{24} b_{24}} R_{b_1 .. b_{24}} \ , \ \eqno(2.8) $$
$$
I_c(R^{a_1 .. a_{23},b}) = + \eta^{a_1 c_1} .. \eta^{a_{23} c_{23}} \eta^{bd} R_{c_1 .. c_{23},d} \ , \ $$
$$
I_c(R^{a_1 .. a_{25},b_1 .. b_{19}}) = + \eta^{a_1 c_1} .. \eta^{a_{25} c_{25}} \eta^{b_1 d_1} .. \eta^{b_{19} d_{19}} R_{c_1 .. c_{23},d_1 .. d_{19}} \ ; \ \ldots $$
\par
The involution invariant sub-algebra $I_c(K_{27})$ is given by 
$$
J_{a_1 a_2} = \eta_{a_1 e} K^e{}_{a_2} - \eta_{a_2 e} K^e{}_{a_1} \  ; \ S_{a_1 a_2} = R^{b_1 b_2} \eta_{b_1 a_1} \eta_{b_2 a_2} - R_{a_1 a_2} \ , \ $$
$$
S_{a_1 .. a_{22}} = R^{b_1 .. b_{22}} \eta_{b_1 a_1} .. \eta_{b_{22} a_{22}} - R_{b_1 .. b_{22}} \ ;  \ S_{a_1 .. a_{24}} = R^{b_1 .. b_{24}} \eta_{b_1 a_1} .. \eta_{b_{24} a_{24}} + R_{b_1 .. b_{24}} \ , \eqno(2.9) $$
$$
S_{a_1 .. a_{23},b} = R^{c_1 .. c_{23},d} \eta_{c_1 a_1} .. \eta_{c_{23} a_{23}} \eta_{db} + R_{a_1 .. a_{23},b} \ ,  \ \ldots $$
$$
S_{a_1 .. a_{25},b_1 .. b_{19}} = R^{c_1 .. c_{25},d_1 .. d_{19}} \eta_{c_1 a_1} .. \eta_{d_{19} b_{19}} + R_{a_1 .. a_{25},b_1 .. b_{19}} \ ; \ \ldots $$
\par 
The algebra of $I_c(K_{27})$ involving generators of $K_{27}$ up to levels $\pm 1$ is given as
$$
[J_{a_1 a_2},J_{b_1 b_2}] =  4 \eta_{[a_1|[b_1} J_{b_2]|a_2]} \ \ , \ \ [J_{a_1 a_2},S_{b_1 b_2}] = 4 \eta_{[a_1|[b_1} S_{b_2]|a_2]} \ \ , $$
$$
[J_{a_1 a_2},S_{b_1 .. b_{22}}] = 2 \cdot 22 \eta_{[a_1|[b_1} S_{b_2 .. b_{22}]|a_2]} \ , \  [J_{a_1 a_2},S_{b_1 .. b_{24}}] = 2 \cdot 24 \eta_{[a_1|[b_1} S_{b_2 .. b_{24}]|a_2]} \ ,  \eqno(2.10) $$
$$
[S_{a_1 a_2},S_{b_1 b_2}] = + 4 \eta_{[a_1|[b_1} J_{b_2]|a_2]} \ \ , \ \ [S_{a_1 a_2},S_{b_1 .. b_{22}}] = S_{a_1 a_2 b_1 .. b_{22}} + S_{b_1 .. b_{22} [a_1,a_2]} \ , $$
$$
[S^{a_1 .. a_{22}},S_{b_1 .. b_{22}}] = S^{a_1 .. a_{22}}{}_{[b_1 b_2 b_3,b_4 .. b_{22}]} - (22)^2 21! \delta^{[a_1 .. a_{21}}_{[b_1 .. b_{21}} J^{a_{22}]}{}_{b_{22]}}  \ .$$
In Appendix A we have given the algebra of $I_c(K_{27})$ involving generators of $K_{27}$ up to levels $\pm 2$.
\par 
The $l_1$ representation of $K_{27}$ is its first fundamental representation, sometimes referred to as the vector representation. It members can also be classified in terms of levels $(l_{26},l_{27})$. The number of up minus down indices on a generator is equal to $22 l_{26} + 2 l_{27} - 1$. The $l_1$ representation up to level two  is given by
$$
P_a \ (0,0) \ ; \ Z^{a} \ (0,1) \ , \ Z^{a_1 .. a_{21}} \ (1,0) \ ; \ Z_{\{1\}}^{a_1 .. a_{23}} \ (1,1) \ ,  \ Z_{\{2\}}^{a_1 .. a_{23}} \ (1,1) \ , $$
$$
Z^{a_1 .. a_{22},b} \ (1,1) \ , \ Z^{a_1 .. a_{24},b_1 .. b_{19}} \ (2,0) \ , \ Z^{a_1 .. a_{25},b_1 .. b_{18}} \ (2,0) \ ; \ \ldots
\eqno(2.11)$$ 
Here we have again separated the generators of different levels  by semi-colons, the indices are anti-symmetric in each block, and we have placed subscripts in curly brackets to distinguish generators of non-zero multiplicity. The listed generators satisfy the following irreducibility conditions
$$
Z^{[a_1 .. a_{22},b]}  = Z^{[a_1 .. a_{24},b_1] b_2 .. b_{19}} = Z^{[a_1 .. a_{25},b_1]b_2 .. b_{18}} = \ldots = 0  
\eqno(2.12)$$
where the $\ldots$ represent similar irreducibility conditions on the additional mixed symmetry tensors above level two in equation (2.11).
\par 
The commutators  of $K_{27}$ with the $l_1$ representation with  generators up to levels $\pm 1$  in $K_{27}$ and up to level one in the $l_1$ representation are  given by
$$
[K^a{}_b,P_c] = - \delta^a{}_c P_b + {1 \over 2} \delta^a{}_b P_c \ , \ [K^a{}_b,Z^{c}] = \delta^{c}{}_b Z^{a} + {1 \over 2} \delta^a{}_b Z^c \ , \  $$
$$
[K^a{}_b,Z^{c_1 .. c_{21}}] = 21 \delta^{[c_1}{}_b Z^{|a|c_2 .. c_{21}]} + {1 \over 2} \delta^a{}_b Z^{c_1 .. c_{22}}  \ , 
[R,P_a] = 0 \ \ , 
\ \ [R,Z^{a}] = Z^a \ \ ,
$$
$$
 \ \ [R,Z^{a_1 .. a_{21}}] = - Z^{a_1 .. a_{21}} \ \ , 
[R^{a_1 a_2},P_b] = 2 \delta^{[a_1}{}_b Z^{a_2]} \ , 
\ [R^{a_1 .. a_{22}},P_b] = 22 \delta^{[a_1}{}_b Z^{a_2 .. a_{22}]}  \ ,  $$
$$
[R^{a_1 a_2},Z^{b}] = 0 \ ,    \ , 
[R^{a_1 .. a_{22}},Z^{b_1 .. b_{21}}] = Z^{a_1 .. a_{22}[b_1 b_2,b_3 .. b_{21}]} + Z^{a_1 .. a_{22} [b_1 b_2 b_3,b_4 .. b_{21}]} \ , $$
$$
[R_{a_1 a_2},P_b] = 0 \ , \ [R_{a_1 .. a_{22}},P_b] = 0 \ , \ [R_{a_1 a_2},Z^b] = - 10 \delta^b_{[a_1} P_{a_2]} \ , \ [R_{a_1 .. a_{22}},Z^b] = 0 \ , \  $$
$$
[R_{a_1 a_2},Z^{b_1 .. b_{21}}] = 0 \ , \ [R_{a_1 .. a_{22}},Z^{b_1 .. b_{21}}] = - 11 \cdot 21! \delta_{[a_1 .. a_{21}}^{b_1 .. b_{21}} P_{a_{22}]} \ . \eqno(2.13)
$$
where the constants $c_1$ and $c_2$ have yet to be determined. 
\par
The commutators  of $I_c(K_{27})$ with the $l_1$ representation, up to level one in $I_c(K_{27})$ and level one in the $l_1$ representation are  given by
$$
[J_{a_1 a_2},P_b] = - 2 \eta_{b [a_1} P_{a_2]} \ , \  [J^{a_1 a_2},Z^{b}] = 2 \eta^{b[a_1} Z^{a_2]}  \ , $$
$$
[J_{a_1 a_2},Z^{b_1 .. b_{21}}] = 2 \cdot 21 \eta_{e[a_1}  \delta^{[b_1}{}_{|a_2]} Z^{|e|b_2 .. b_{21}]} \ , 
[S^{a_1 a_2},P_b] =  2 \delta^{[a_1}{}_b Z^{a_2]}  \ , \
$$
$$
[S_{a_1 a_2},Z^{b}] = + 10 \delta^b_{[a_1} P_{a_2]} \  , \ 
[S^{a_1 .. a_{22}},P_b] = 22 \delta^{[a_1}{}_b Z^{a_2 .. a_{22}]} \ , $$
$$
[S_{a_1 .. a_{22}},Z^{b_1 .. b_{21}}] = + 11 \cdot 21!  \delta_{[a_1 .. a_{21}}^{b_1 .. b_{21}} P_{a_{22}]} + Z_{a_1 .. a_{22}}{}^{[b_1 b_2,b_3 .. b_{21}]} + Z_{a_1 .. a_{22}}{}^{[b_1 b_2 b_3,b_4 .. b_{21}]}  \ . 
 \eqno(2.14)$$


\medskip
{{\bf 3. Non-linear realisation of $K_{27} \otimes_s l_1$}}
\medskip
We now consider the nonlinear realization of $K_{27} \otimes_s l_1$ with  local subgroup $I_c(K_{27})$. The method is essentially  the same as that for $E_{11}$  [6,8,9,10]. We begin by considering a group element of $K_{27} \otimes_s l_1$ which can be taken in the form
$$
g(x) = g_l(x) g_K(x) \eqno(3.1)$$
Here the element $g_l$ may be taken as a product of exponentials of the form $\Pi_A e^{x^A l_A}$, where the $l_A$ are the generators of the $l_1$ representation, and the $x^A$ are interpreted as the coordinates of a generalized spacetime associated to the $l_1$ representation. The element $g_K$ in $K_{27}$ can be taken as a product of exponentials of the form $\Pi_{\alpha} e^{A_{\alpha} R^{\alpha}}$, where $R^{\alpha}$ are the generators of $K_{27}$. The fields $A_{\alpha}$ are taken to depend on the generalized space-time coordinates $x^A$.
\par 
The non-linear realisation is, by definition, invariant under the following transformations
$$
g \ \to \ g_0 g \ , \ g_0 \in K_{27} \otimes_s l_1 \ , \ {\rm along \ with \ } \ g \to g h \ , \ h \in I_c(K_{27}) \ . 
\eqno(3.2)$$
Here the element $g_0$ is taken to be a rigid transformation, meaning it is independent of the space-time coordinates $x^A$, while $h$, belonging to $I_c(K_{27})$,  and it is taken to be a local transformation and so depends on $x$.
\par 
Since  $l_1$ is a representation of $K_{27}$, equation (3.2) can equivalently be written using equation (3.1) as 
$$
g_l \to g_0 g_l g_0^{-1} \ , \ g_K \to g_0 g_K \ , \ g_K \to g_K h \ . 
\eqno(3.3)$$
\par
This invariance under local $I_c(K_{27})$ transformations allows us to choose a gauge in which the group element $g_K$ depends only on the non-negative generators from $K_{27}$ as in equation (3.1). We will take this choice and it will be essential to preserve this gauge choice throughout what follows.
\par 
Using the local $I_c(K_{27})$ transformations, we thus parametrize our group element in terms of non-negative generators as
$$
g_K(x) = \ldots e^{A_{a_1 .. a_{25},b_1 .. b_{19}} R^{a_1 .. a_{25},b_1 .. b_{19}}}  e^{h_{a_1 .. a_{23},b} R^{a_1 .. a_{23},b}}  e^{A_{a_1 .. a_{24}} R^{a_1 .. a_{24}}}  $$
$$
 e^{A_{a_1 .. a_{22}} R^{a_1 .. a_{22}}}  \ e^{A_{a_1 a_2} R^{a_1 a_2}} e^{\phi R}  e^{h_a{}^b K^a{}_b} 
 \eqno(3.4)$$
where the $\ldots$ represents contributions from generators of $K_{27}$ above level two. The coefficients of the generators will turn out to be the fields, which we list 
$$
h_a{}^b \ , \ \phi \ ; \ A_{a_1 a_2} \ , \ A_{a_1 .. a_{22}} \ ; \ A_{a_1 .. a_{24}} \ , \ h_{a_1 .. a_{23},b} \ , \ A_{a_1 .. a_{25},b_1 .. b_{19}} \ ; \ \ldots 
\eqno(3.5) $$
Here all fields at different levels are separated by a semi-colon, and they possess the same symmetries and irreducibility conditions as their corresponding $K_{27}$ generators, for example blocks of  indices are anti-symmetric. In the fields listed above these irreducibility conditions read as
$$
h_{[a_1 .. a_{23},b]} = A_{[a_1 .. a_{25},b_1] b_2 .. b_{19}} = 0 \ , \eqno(3.6)$$
along with similar irreducibility conditions on the fields above level two
\par 
The element $g_l$ can be written in the form
$$
g_l(x) = ... e^{x_{a_1 .. a_{21}} Z^{a_1 .. a_{21}}}   e^{y_a Z^a} e^{x^a P_a}
 \eqno(3.7)$$ 
The coefficients of these generators are the  coordinates of a generalized space-time
$$
x^a \ ; \ y_{a} \ , \ x_{a_1 \ldots a_{21}} \ ; \ \ldots \ , 
\eqno(3.8) $$
which also possess the same index symmetries and irreducibility conditions as the corresponding $l_1$ generators.
\par 
We will find out that the field $h_a{}^b$ is the graviton field, $\phi$ is the dilaton field, $A_{a_1 a_2}$ is the Kalb-Ramond field, $A_{a_1 .. a_{22}}$ is the dual Kalb-Ramond field, $A_{a_1 .. a_{24}}$ is the dual dilaton field and $h_{a_1 .. a_{23},b}$ is the dual graviton field. The coordinates $x^a$ are the usual coordinates of 26-dimensional space-time. 
\par 
The dynamics of the non-linear realisation is a set of equations that transform covariantly under equations (3.2), or equivalently equation (3.3). We will find  these equations at low levels in Sections 4 and 5. The simplest way to find these equations is to construct them from the  Maurer-Cartan forms
$$
{\cal V} = g^{-1} d g = {\cal V}_l + {\cal V}_K
 \eqno(3.9)$$
where  the Cartan forms ${\cal V}_l$ contain the $l_1$ generators and ${\cal V}_K$ the  $K_{27}$ generators. We can write 
$$
{\cal V}_l = g_K^{-1} (g_l^{-1} d g_l) g_K \equiv dx^{\Pi} E_{\Pi}{}^A l_A  \ \, \ \ {\cal V}_K \equiv g_K^{-1} d g_K = dx^{\Pi}  G_{\Pi,\alpha} R^{\alpha} \ \ .
 \eqno(3.10)$$
The first part of equation (3.10) is the defining equation for a generalized vielbein $E_{\Pi}{}^A$ which transforms on its $A$ index by  a  local tangent space $I_c(K_{27})$ transformation and on its $\Pi$ index by a $K_{27}\otimes_s l_1$  transformation which acts on   the generalized space-time. The Cartan form coefficients $G_{\Pi,\alpha}$ contain the  fields in equation (3.5). Using the vielbein we can re-write the Cartan as
$$
{\cal V} = dx^{\Pi} E_{\Pi}{}^A (l_A + G_{A,\alpha} R^{\alpha})
 \eqno(3.11) $$
which is now expressed in terms of the Cartan form coefficients $G_{A,\alpha} = (E^{-1})_A{}^{\Pi} G_{\Pi,\alpha}$ which only carries tangent indices space and so it only transforms under the tangent space $I_c(K_{27})$  transformations.
\par 
The vielbein in ${\cal V}_l$ of equation (3.9) can be computed from its definition in equation (3.10) and is given by at low levels by 
$$
E_{\Pi}{}^A = (\det e)^{-{1 \over 2}} \pmatrix{ e_{\mu}{}^a & - 2 e^{- \phi} A_{\mu a}   & - 22 e^{ \phi}  A_{\mu a_1 .. a_{21}}  & \dots \cr
0 & e^{-  \phi} e^{\mu}_a & 0 & \dots \cr
0 & 0 & e^{ \phi} e^{\mu_1 .. \mu_{21}}_{a_1 .. a_{21}}  & \dots \cr
\vdots & \vdots & \vdots & \ddots} 
\eqno(3.12)$$
\par 
We may write the Cartan form ${\cal V}_K$ of equation (3.9) as 
$$
{\cal V}_K = G_a{}^b K^a{}_b + G R + G_{a_1 a_2} R^{a_1 a_2} + G_{a_1 .. a_{22}} R^{a_1 .. a_{22}} + G_{a_1 .. a_{24}} R^{a_1 .. a_{24}} $$
$$
+ G_{a_1 .. a_{23},b} R^{a_1 .. a_{23},b} + G_{a_1 .. a_{25},b_1 .. b_{19}} R^{a_1 .. a_{25},b_1 .. b_{19}} + \ldots 
\eqno(3.13)$$
These can be  computed by inserting $g_K$  of  equation (3.4) into ${\cal V}_K $ to find that 
$$
G_a{}^b = (e^{-1} d e)_a{}^b \ \ , G = d \phi \ \ ; \ \ G_{a_1 a_2} = e^{- \phi}  e_{\mu_1}{}^{a_1} e_{a_2}{}^{\mu_2} d A_{\mu_1 \mu_2} \ ,  $$
$$
G_{a_1 .. a_{22}} = e^{+ \phi}  e_{a_1}{}^{\mu_1} \ldots e_{a_{22}}{}^{\mu_{22}} d A_{\mu_1 .. \mu_{22}} \ \ ; $$
$$
G_{a_1 .. a_{24}} =  e_{a_1}{}^{\mu_1} \ldots e_{a_{24}}{}^{\mu_{24}} (dA_{\mu_1 .. \mu_{24}} - d A_{[\mu_1 .. \mu_{22}} A_{\mu_{23} \mu_{24}]}) \ \ , \eqno(3.14)
 $$
$$
G_{a_1 .. a_{23},b} = e_{a_1}{}^{\mu_1} \ldots e_{a_{23}}{}^{\mu_{23}} e_b{}^{\nu} (d h_{\mu_1 .. \mu_{23},\nu} - d A_{[\mu_1 .. \mu_{22}} A_{\mu_{23}]\nu} +  d A_{[\mu_1 .. a_{22}} A_{\mu_{23}\nu]} ) \ \ , $$
$$
G_{a_1 .. a_{25},b_1 .. b_{19}} = e^{+2 \phi} e_{a_1}{}^{\mu_1} \ldots  e_{b_{19}}{}^{\nu_{19}} ( d A_{\mu_1 .. \mu_{25},\nu_1 .. \nu_{19}}  - {1 \over 2} A_{[\mu_1 .. \mu_{22}} d A_{\mu_{23} \mu_{24} \mu_{25} ] \nu_1 .. \nu_{19}}  -{\rm irred} ) \ ; $$
where $e_{\mu}{}^{a} \equiv (e^h)_{\mu}{}^a$. We have written the above using differential forms $d = dx^{\Pi} \partial_{\Pi} $. In the $G_{25,19}$ Cartan form, the `$-{\rm irred}$' term signifies that we should subtract off whatever is needed to ensure that the Cartan form coefficient is irreducible as in equation (2.3).
\par 
We now consider how the Cartan form ${\cal V}$ in equation (3.9) transforms under the non-linear realisation transformations of equation (3.2). Under the rigid transformations, ${\cal V}$ remains invariant, however under the local $I_c(K_{27})$ tranformations we have
$$
{\cal V}_l \ \to \  h^{-1} {\cal V}_l h \ \ , \ \ {\cal V}_K \ \to \  h^{-1} {\cal V}_K h + h^{-1} dh \ \ . 
\eqno(3.15)$$
 The local transformation $h$ can be written as $h = I - \Lambda^{\alpha} S_{\alpha}$, where $S_{\alpha}\in I_c(K_{27})$.  
\par 
We first focus on the transformation of ${\cal V}_K$ in equation (3.15). Infinitesimally, this reads as 
$$
\delta G_{\alpha} R^{\alpha} = [\Lambda^{\alpha}S_{\alpha},G_{\beta} R^{\beta}] - d\Lambda^{\alpha} S_{\alpha} \ \ 
 \eqno(3.16)$$
where the $R^{\alpha}$ are the non-negative generators of $K_{27}$.
We will compute the transformations up to and including level one
and so we will take  $\Lambda^{\alpha} S_{\alpha} = \Lambda^{a_1 a_2} S_{a_1 a_2} + \Lambda^{a_1 .. a_{22}} S_{a_1 .. a_{22}}$. 
\par 
Explicitly, the variations $\delta G_{\alpha}$ in equation (3.16) are given by
$$
\delta G_a{}^b = 4 \Lambda^{eb} G_{ea} - {1 \over 6} \Lambda^{e_1 e_2} G_{e_1 e_2} \delta^b{}_a + (22)^2 21! \Lambda^{e_1 .. e_{21} b} G_{e_1 .. e_{21}a}  - {11 \over 12} 22! \Lambda^{e_1 .. e_{22}} G_{e_1 .. e_{22}} \delta^b{}_a \ , $$
$$
\delta G = {1 \over 3} \Lambda^{e_1 e_2} G_{e_1 e_2} - {22! \over 6} \Lambda^{e_1 .. e_{22}} G_{e_1 .. e_{22}} \ ;  $$
$$
\delta G_{a_1 a_2} = - d \Lambda_{a_1 a_2} - \Lambda_{a_1 a_2} G + 2 \Lambda_{e [a_1} G_{a_2]}{}^e + {24! \over 12}  \Lambda^{e_1 .. e_{22}} G_{e_1 .. e_{22} a_1 a_2}  $$
$$
 + {23! \cdot 11 \over 6} \Lambda^{e_1 .. e_{22}} G_{e_1 .. e_{22}[a_1,a_2]}   -  {23! \cdot 11 \over 6} \Lambda^{e_1 .. e_{21} e_{22}} G_{a_1 a_2 [e_1 .. e_{21},e_{22}]} \ ,  \eqno(3.17)
$$
$$
\delta G_{a_1 .. a_{22}} = - d \Lambda_{a_1 .. a_{22}} + \Lambda_{a_1 .. a_{22}} G + 22 \Lambda_{e[a_1  .. a_{21}} G_{a_{22}]}{}^e  - 4 \cdot 23 \Lambda^{e_1 e_2} G_{e_1 e_2 a_1 .. a_{22}} $$
$$
 - {23 \cdot 11 \over 3}  \Lambda^{e_1 e_2} G_{a_1 .. a_{22}[e_1,e_2]}   + {23 \cdot 11 \over 3} \Lambda^{e_1 e_2} G_{e_1 e_2 [a_1 .. a_{21},a_{22}]}   +\ldots  $$
$$
\delta G_{a_1 .. a_{24}}  = \Lambda_{[a_1 a_2} G_{a_3 .. a_{24}]} - \Lambda_{[a_1 .. a_{22}} G_{a_{23} a_{24}]} + \ldots \ , $$
$$
\delta G_{a_1 .. a_{23},b} = G_{[a_1 .. a_{22}} \Lambda_{a_{23}] b} -  G_{[a_1 .. a_{22}} \Lambda_{a_{23} b]} -  \Lambda_{[a_1 .. a_{22}} G_{a_{23}] b} + \Lambda_{[a_1 .. a_{22}} G_{a_{23} b]} +\ldots \ , $$ 
   The $+\ldots$ denote terms of level three and terms involving the Cartan form  
 $ G_{b, a_1\ldots  a_{25}, c_1\ldots c_{19}} $ which we will not need in this paper. We have also not given the transformation of this field. 
\par
Since the $ I_c(K_{27})$ transformation we are carrying out involves the negative level generators $R_{a_1 a_2} $ and 
$R_{a_1 .. a_{22}}$, the transformation of equation (3.16) does not automatically preserve the gauge choice of equation (3.4), which excluded such  negative level generators from the group element. To preserve the gauge we must choose compensating local transformations $ \Lambda^{\alpha}$ so as to  preserve the choice of equation (3.4) and so we take the conditions 
$$
d \Lambda^{a_1 a_2} + 2 \Lambda^{e [a_1} G_e{}^{a_2]} -  \Lambda^{a_1 a_2} G   = 0 \ \ , \eqno(3.18) $$
$$
d \Lambda^{a_1 .. a_{22}} + 22 \Lambda^{e [a_1 .. a_{21}} G_e{}^{a_{22}]}  +  \Lambda^{a_1 .. a_{22}} G    = 0 \ \ . 
\eqno(3.19) $$
\par 
Inserting these  into the variations of equation (3.17) we find that 
$$
\delta G_a{}^b = 4 \Lambda^{eb} \overline{G}_{ea} - {1 \over 6} \Lambda^{e_1 e_2} \overline{G}_{e_1 e_2} \delta^b{}_a + (22)^2 21! \Lambda^{e_1 .. e_{21} b} G_{e_1 .. e_{21}a}  - {11 \over 12} 22! \Lambda^{e_1 .. e_{22}} G_{e_1 .. e_{22}} \delta^b{}_a \ , $$
$$
\delta G = {1 \over 3} \Lambda^{e_1 e_2} \overline{G}_{e_1 e_2} - {22! \over 6} \Lambda^{e_1 .. e_{22}} G_{e_1 .. e_{22}}  \ ; $$
$$
\delta \overline{G}_{a_1 a_2} = 2 \cdot 2 \Lambda_{e [a_1} G_{(a_2]}{}^{e)} - 2 \Lambda_{a_1 a_2} G  + {24! \over 12}  \Lambda^{e_1 .. e_{22}} G_{e_1 .. e_{22} a_1 a_2} \   $$
$$
 + 23! \cdot 2  \Lambda^{e_1 .. e_{22}} G_{e_1 .. e_{22}[a_1,a_2]}     \ , \eqno(3.20) $$
$$
\delta G_{a_1 .. a_{22}} = + 2 \cdot 22 \Lambda_{e[a_1  .. a_{21}} G_{(a_{22}]}{}^{e)} + 2  \Lambda_{a_1 .. a_{22}} G  - 4 \cdot 23 \Lambda^{e_1 e_2} G_{e_1 e_2 a_1 .. a_{22}} 
$$
$$
 - 23 \cdot 4 \Lambda^{e_1 e_2} G_{a_1 .. a_{22}[e_1,e_2]}  +\ldots   $$
$$
\delta G_{a_1 .. a_{24}}  = \Lambda_{[a_1 a_2} G_{a_3 .. a_{24}]} - \Lambda_{[a_1 .. a_{22}} G_{a_{23} a_{24}]}+\ldots  \ , $$
$$
\delta G_{a_1 .. a_{23},b} = G_{[a_1 .. a_{22}} \Lambda_{a_{23}] b} -  G_{[a_1 .. a_{22}} \Lambda_{a_{23} b]} -  \Lambda_{[a_1 .. a_{22}} G_{a_{23}] b} + \Lambda_{[a_1 .. a_{22}} G_{a_{23} b]} 
$$
\par 
We now focus on the transformation of ${\cal V}_l$ in equation (3.15). Since the $l_A$ transform as a representation of $K_{27}$, we have $[R^{\alpha},l_A] = -(D^{\alpha})_A{}^B l_B$. Under a local  $h$ transformations the veirbein transforms as  
$$
\delta E_\Pi{}^A = E_\Pi{}^B \Lambda^{\alpha} (D_{\alpha})_B{}^A 
\eqno(3.21)$$
By evaluating equation (3.15), or (3.21) directly,  we find the vielbein varies at lowest order as 
$$
\delta E_{\Pi}{}^a = 10 \Lambda^{b a} E_{\Pi,b} +  11 \cdot 21! \Lambda^{b_1 .. b_{21} a} E_{\Pi,b_1 .. b_{21}} \ , \  
\eqno(3.22) $$
$$
\delta E_{\Pi}{}_a = 2 \Lambda_{b a} E_{\Pi}{}^b \ \ , \ \ \delta E_{\Pi}{}_{a_1 .. a_{21}} =  22 \Lambda_{b a_1 .. a_{21}} E_{\Pi}{}^b \ .  \ \eqno(3.23)$$
We note that these transformations mix world indices on $ E_{\Pi}{}_a $ corresponding to the usual coordinates of spacetime with the higher level coordinates and vice-versa. 
 We can compute the variation of the inverse vielbein from $ \delta (E^{-1})_{A}{}^{\Pi} = - (E^{-1})_{A}{}^{\Pi'} (\delta E)_{\Pi'}{}^B (E^{-1})_B{}^{\Pi}$. 
\par
In the variation of the Cartan forms under local $ I_c(K_{27})$ of equation (3.20) we took them to carry a world index, that is, $G_{\Pi, \alpha}$. However, when constructing the dynamics we will work with $ G_{A,\alpha} = (E^{-1})_A{}^{\Pi} G_{\Pi,\alpha}$. As such we must include the transformation of the vierbein, indeed 
$$
\delta G_{A,\alpha} = \delta [(E^{-1})_A{}^{\Pi} G_{\Pi,\alpha}] = (\delta E_A{}^{\Pi}) G_{\Pi,\alpha} + E_A{}^{\Pi}  \delta G_{\Pi,\alpha}  \eqno(3.24) $$
where the second term transforms as in equation (3.20), and the first term transforms as the inverse of the vierbein given in  equation (3.21). We find that the derivative tangent space  indices of the Cartan forms transform as
$$
\delta G_{a,\alpha} = - 2 \Lambda_{ae} \hat{G}^{e,}{}_{\alpha} - 22 \Lambda_{a e_1 .. e_{21}} \hat{G}^{e_1 .. e_{21},}{}_{\alpha} \ , \  \delta \hat{G}^{a,}{}_{\alpha} = - 10 \Lambda^{ae} G_{e,\alpha} \ , \
 \eqno(3.25) $$
$$
\delta \hat{G}^{a_1 .. a_{21},}{}_{\alpha} = - 11 \cdot 21! \Lambda^{a_1 .. a_{21}e} G_{e,\alpha} \ . \ 
\eqno(3.26) $$
We will denote a Cartan form that has a spacetime derivative with respect to a higher level coordinate by putting a hat on the Cartan form, for example $G_{a,\alpha}$ and $\hat G_{a,\alpha}$ have derivatives with respect to $x^\mu$ and $y_\mu$ respectively. 
Of course the full variation of $\delta G_{A,\alpha} $ is the sum of these last two equations and that of equation (3.20).

\medskip
{{\bf 4. First order duality relations}}
\medskip
The non-linear realisation contains the graviton $h_a{}^b$, the scalar $\phi$ and the two form $A_{a_1a_2}$, but it also contains their dual fields. At the lowest level these are   the fields $h_{a_1\ldots a_{23} ,b}$,  $\phi_{a_1\ldots a_{24}}$ and   $A_{a_1\ldots a_{22}}$ respectively. In this section we will use the symmetries of the non-linear realisation to find the corresponding duality relations. More precisely we will find a set of duality relations that transform into each other under the symmetries of the non-linear realisation. We will construct these duality relations out of the Cartan forms of  equation (3.14) which only transform under $I_c(K_{27})$ transformations. These relations are first order in derivatives,  and as explained in [9,10,8], we will construct them up to, and including,  derivatives with respect to the level one coordinates. 
\medskip
{{\bf 4.0 Summary of duality relations}}
\medskip¤
For ease of access we summarise the duality relations we will  find in this section:
$$
D_a \equiv G_a + e_1 \ \varepsilon_a{}^{b_1 .. b_{25}} G_{b_1,b_2 .. b_{25}} = 0 \ \ , \ \ e_1 = {1 \over 12} \ \ , \eqno(4.0.1)$$
$$
\overline{D}_{a_1 a_2 a_3} \equiv \overline{G}_{[a_1, a_2 a_3]} + e_2 \ \varepsilon_{a_1 a_2 a_3}{}^{e_1 .. e_{23}} G_{e_1,e_2 .. e_{23}} = 0 \ \ , \ \ e_2 = {1 \over 6} \ \ ,  \eqno(4.0.2)$$
$$
D_{a,b_1 b_2} \equiv  (\det e)^{{1 \over 2}} \omega_{a,b_1 b_2} + e_3 \ \varepsilon_{b_1 b_2}{}^{e_1 .. e_{24}} G_{e_1,e_2 .. e_{24},a}  \dot = 0 \ \ , \ \ e_3 = - 1 \ \ . 
\eqno(4.0.3)$$
We have only kept terms which   contain derivatives with respect to the usual (level 0) coordinates of spacetime. We give  the values of the coefficients $e_1,e_2,e_3$ which are uniquely fixed (up to an overall $\pm$ sign) by the symmetries of the non-linear realisation. In these equations we defined $G_{a,b_1 b_2} = G_{a,b_1}{}^e \eta_{e b_2}$ and the spin connection by 
$$
(\det e)^{{1 \over 2}} \omega_{a,b_1 b_2} \equiv  - G_{b_1,(b_2 a)} + G_{b_2,(b_1 a)} + G_{a,[b_1 b_2]} .
 \eqno(4.0.4) $$
\par
As we derive  our results we will encounter the same duality relations but with the epsilon symbol acting on the other dual  field and so for convenience we list these relations 
$$
\eqalign{
{D}_{a_1 a_2 \ldots  a_{23}} &\equiv  G_{[a_1 , a_2 \ldots  a_{23}]} + {1 \over 3! 23! e_2} \varepsilon_{a_1 a_2\ldots a_{23}}{}^{e_1 e_2 e_3} \overline{G}_{e_1,e_2 e_3} \cr
&\equiv {1 \over 3! 23! e_2} \varepsilon_{a_1 a_2\ldots a_{23}}{}^{e_1 e_2 e_3} \overline{D}_{e_1 e_2 e_3} \ , }
$$
$$
D_{a_1 a_2 \ldots _{25}} \equiv  G_{[a_1 a_2 \ldots _{25}]}+{1 \over 25! e_1} \varepsilon^{a_1a_2 \dots a_{25} e} G_e={1 \over 25! e_1} \varepsilon^{a_1a_2 \dots a_{25} e} D_e \ ,
$$
$$
\eqalign{
D_{a_1 a_2 \ldots a_{24},b} &\equiv G_{[a_1 , a_2 \ldots a_{24}],b}- {1 \over 2! 24! e_3} \varepsilon_{a_1 a_2\ldots  a_{24}}{}^{e_1 e_2} (\det e)^{{1 \over 2}} \omega_{b,e_1 e_2} \cr
&= - {1 \over 2! 24! e_3} \varepsilon_{a_1 a_2\ldots  a_{24}}{}^{e_1 e_2} D_{b ,  e_1 e_2} \ , }
 \eqno(4.0.5)$$

\medskip
{{\bf 4.1 Kalb-Ramond duality}}
\medskip
We now establish the duality relation between the Kalb-Ramond field $A_{a_1 a_2}$ at level one, and the lowest level field that can possibly be its dual, the  field $A_{a_1 \ldots a_{22}}$. Since the duality relation is first order in spacetime derivatives it must be a relation between the corresponding Cartan forms $\bar G_{b, a_1 a_2}$ and $G_{b, a_1 \ldots a_{22}}$. 
On grounds of Lorentz symmetry it must be of the form
$$
\overline{D}_{a_1 a_2 a_3} \equiv \overline{G}_{[a_1,a_2 a_3]} + e_2 \varepsilon_{a_1 a_2 a_3}{}^{e_1 .. e_{23}} G_{e_1,e_2 .. e_{23}} = 0. \eqno(4.1.1)$$
The coefficient in equation (4.1.1) will be  fixed uniquely (up to an overall sign) by considering the variation of this duality relation under the $I_c(K_{27})$ transformations of equation (3.20). 
\par 
Examining the  $I_c(K_{27})$ variation of $\bar G_{[ a_1, a_2 a_3]}$ of equation (3.20),  we find that it varies into $G_{a_1, a_2 \ldots a_{25}}$ but in order for the duality relation to hold it must vary into  terms which contain $G_{ [a_1, a_2 \ldots a_{25}]}$, that is, the index on the spacetime derivative ($l_1$) is anti-symmetrized with $K_{27}$ indices on the fields.   This difficulty can be overcome if we add certain  terms ( $l_1$ terms) which involve derivatives with respect to the level one coordinates to $\bar G_{[ b, a_1 a_2]}$ . As such we introduce the $l_1$-extension of  $\bar {G}_{[ b, a_1 a_2]}$ denoted by $\bar {\cal G}_{[ b, a_1 a_2]}$:
 $$
\eqalign{
\overline{{\cal G}}_{[a_1 a_2 a_3]} \equiv \overline{G}_{[a_1 , a_2 a_3]} &+ {1 \over 5} \hat{G}_{[a_1 ,a_2 a_3]} + {4 \cdot 22 \cdot 23\over 3} \hat{G}^{e_1 .. e_{21},}{}_{e_1 .. e_{21} a_1 a_2 a_3} \cr
&+   2 \cdot 22 \cdot 23 \hat{G}^{e_1 .. e_{21},}{}_{e_1 .. e_{21} [a_1 a_2,a_3]} } 
\eqno(4.1.2)$$
Its variation under the $I_c(K_{27})$ transformations of equation (3.20) is given by
$$ 
\delta \overline{{\cal G}}_{[a_1a_2 a_3]} = - 2  \Lambda_{[a_1 a_2}G_{[a_3]} + {25!\over 12.3} \Lambda^{e_1\ldots e_{22}} G_{[e_1, e_2 \ldots e_{22} a_1a_2a_3] } 
$$
$$
 + 2 \Lambda^e{}_{[a_1} (\det e)^{{1 \over 2}} \omega_{|e|,a_2 a_3]} 
+ 24! \Lambda^{e_1 .. e_{22}} G_{[e_1,e_2 .. e_{22} [a_1 a_2],a_3]}
 \eqno(4.1.3)$$
\par 
Following the same argument we must replace the second term  $G_{[e_1,e_2 .. e_{23}]} $  such that its variation contains anti-symmetrized indices by adding $l_1$ terms. Using equations (3.22) and (3.23) we find that the  $l_1$-extension of the anti-symmetrized Cartan form is given by
$$
{\cal G}_{[a_1 a_2 .. a_{23}]}  \equiv  G_{[a_1,a_2 \ldots  a_{23}]}  + {2 \over  21!}  \hat{G}_{[a_1\ldots   a_{21},a_{22} a_{23}]} -  {4  \over 5}\hat{G}^{c,}{}_{c a_1 .. a_{23}} +  {2  \over 5} \hat{ G}^{c,}{}_{a_1 .. a_{23},c}  
\eqno(4.1.4)$$
Its  variation is given by 
$$
\delta {\cal G}_{[a_1,a_2 .. a_{23}]} =  22  \Lambda^{e}{}_{ [a_1 .. a_{21}} (\det e)^{{1 \over 2}} \omega_{|e|,a_{22} a_{23}]} + 2 \Lambda_{[a_1a_2 \ldots a_{22}} G_{a_{23}]}
$$
$$
 -4.25 \Lambda^{e_1e_2} G_{[a_1,a_2\ldots a_{23}e_1e_2]} -4.24  \Lambda^{e_1e_2} G_{[a_1,a_2\ldots a_{23} e_1], e_2}
\eqno(4.1.5)$$
\par
Taking the above discussion on the addition of $l_1$ terms we now replace equation (4.1.1) to have the duality relation
 $$
\overline{\cal D}_{a_1a_2 a_3} \equiv \overline{{\cal G}}_{[a_1a_2 a_3]}  + e_2 \varepsilon_{a_1 a_2 a_3}{}^{e_1 .. e_{23}} {\cal G}_{e_1 e_2 .. e_{23}}  \eqno(4.1.6)$$
  In this paper quantities that have $l_1$ terms added are denoted by caligraphic symbols. 
  \par
  Using equations (4.1.3) and (4.1.5) the variation of the duality relation of equation (4.1.6) is given by 
 $$
  \delta \overline{\cal D}_{a_1 a_2 a_3} = -2D_{[a_1} \Lambda_{a_2 a_3]} +2e_2\varepsilon_{a_1 a_2 a_3}{}^{e_1 .. e_{23}} \Lambda_{e_1 .. e_{22}} D_{e_{23}}  
 $$
 $$
 + 2 \Lambda^e{}_{[b_1}  \{D_{|e|,b_2 b_3]}  - (e_3+ 6 e_2 )\varepsilon_{b_2 b_3]}{}^{e_1 \ldots e_{24}} G_{e_1,e_2 \ldots e_{24},e} \}  + 22 e_2 \varepsilon_{a_1 a_2 a_3}{}^{c_1 .. c_{23}} \Lambda^d{}_{c_1 .. c_{21}} D_{d,c_{22} c_{23}}
 $$
 $$
 +25!\Lambda ^e_2\ldots e_{22} G_{[a_1,a_2a_3e_1\ldots e_{22}]}(2e_2e_1-{1\over 36})
+ \Lambda _{[a_1a_2}\epsilon_{a_3]}{}^{c_1 c_2\ldots c_{25}}G_{[c_1,c_2\ldots c_{25]} }(2e_1-e_2)
 $$
 $$
+ 24! (1 + 6 e_2 e_3) \Lambda^{e_1 .. e_{22}} G_{[e_1,e_2 .. e_{22}[a_1 a_2],a_3]} 
  \eqno(4.1.7)$$
It is only consistent to set the duality relation of equation (4.1.6) to zero if it varies into other duality duality relations. 
The first four terms  already appears as a sum of duality relations. The remaining terms  cannot be identified as duality relations and so we must set their coefficients to zero. As such we impose the relations:
$$
{e_2 \over e_1} = 2,\ {1 \over 72 e_1 e_2}=1,\  (1 + 6 e_2 e_3)=0 ,\ 6e_2=-e_3
\eqno(4.1.8)$$
Up to a minus sign, which we choose,  these fix the values of the coefficients to be those in equations (4.01-4.03). The reader who has followed this calculation in detail will observe that it could fail in many palaces but the $K_{27}$ symmetry always ensures that the terms collaborate in just such a way that is works. 
\par 
With these values the $I_c(K_{27})$ variation of the duality relation of  equation (4.1.7) reads as
$$
 \delta \overline{\cal D}_{b_1 b_2 b_3} =  {11 \over 3} \varepsilon_{b_1 b_2 b_3}{}^{e_1 .. e_{23}} \Lambda^c{}_{e_1 .. e_{21}} D_{c ,e_{22} e_{23}} - 2 D_{[b_1} \Lambda_{b_2 b_3]} 
 $$
 $$
+ {1 \over 3 } \varepsilon_{b_1 b_2 b_3}{}^{e_1 .. e_{23}} \Lambda_{e_1 .. e_{22}} D_{e_{23}}  + 2 \Lambda^e{}_{[b_1} D_{|e| , b_2 b_3]}   
\eqno(4.1.9)$$


\medskip
{{\bf 4.2 Gravity duality}}
\medskip
We now consider the duality relation between the graviton at level zero, and the dual graviton at level two. We adopt as our starting point the gravity duality we found in the previous section. 
$$
D_{a, b_1 b_2} \equiv  (\det e)^{{1 \over 2}} \omega_{a,b_1 b_2} + e_3 \varepsilon_{b_1 b_2}{}^{e_1 .. e_{24}} G_{[e_1,e_2 .. e_{24}],a} \dot = 0. \eqno(4.2.1)$$
Here we leave the coefficient $e_3$ arbitrary so that the reader can see the consistency of the calculation.  Lorentz symmetry implies that equation (4.2.1) is indeed the only possible duality relation between the graviton and dual graviton which involves only the usual spacetime derivatives. As explained in [10,19,20] this duality relation must be thought of as an equivalence relation, hence the symbol $\dot =$ instead of the usual equals sign. In fact it holds up to a local Lorentz transformation of the spin connection. 
\par 
As in the previous section we must modify the Cartan forms  by adding $l_1$ terms such  that their variations involve curls in the usual spacetime indices.  We define the following $l_1$-extension of the spin connection  in equation (4.0.4) as follows 
$$
(\det e)^{{1 \over 2}} \Omega_{a,b_1 b_2} \dot{\equiv} (\det e)^{{1 \over 2}} \omega_{a,b_1 b_2} +{2\over 5} \hat{\overline{G}}_{(a|,[b_1|b_2)]} + {1 \over 15} \eta_{(a[b_2)} \hat{\overline{G}}^{e,}{}_{e|b_1]}    
$$
$$
+ 2 \cdot 22 \cdot 21 \hat{G}^{e_2 .. e_{21}}{}_{(a|,e_2 .. e_{21} [b_1 | b_2)]}  + {11\cdot 22 \over 3} \eta_{(a[b_2)|} \hat{G}^{e_2 .. e_{22},}{}_{e_2 .. e_{22}|b_1]}
\eqno(4.2.2)$$
It's variation is given by 
$$
\delta [ (\det e)^{{1 \over 2}} \Omega_{a,b_1 b_2}] \dot = 6 \Lambda^e{}_a \overline{G}_{[b_1,b_2 e]} - 6 \Lambda^e{}_{[b_1} \overline{G}_{[b_2],ae]} - {1 \over 2} \eta_{a[b_1} \overline{G}_{[b_2],e_1 e_2]} \Lambda^{e_1 e_2}  
$$
$$
+11 \cdot 23! \Lambda^{e_1 .. e_{21}}{}_a G_{[e_1,e_2 .. e_{22} b_1 b_2]} + 11 \cdot 23! \Lambda^{e_1 .. e_{21}}{}_{[b_1} G_{[b_2],e_1 .. e_{21}a]} 
$$
$$
- {11 \cdot 23! \over 12} \eta_{a[b_1} G_{[b_2],e_1 .. e_{22}]} \Lambda^{e_1 .. e_{22}} 
 \eqno(4.2.3)$$
We observe that the derivative indices are indeed anti-symmetrized with the indices on the fields. As we will see the second term in the duality relation (4.2.1) does not require an $l_1$ extension for the case of interest here. 
\par 
We thus replace the duality relation of equation (4.2.1) by it's $l_1$-extended version 
$$
{\cal D}_{a,b_1 b_2} \equiv (\det e)^{{1 \over 2}} \Omega_{a,b_1 b_2}+ e_3 \varepsilon_{b_1 b_2}{}^{e_1 .. e_{24}} G_{e_1,e_2 .. e_{24},a} 
\eqno(4.2.4)$$
\par
As a first step in varying this duality relation we vary the second term and express some of the terms in duality relations to find that 
$$
\delta [e_3 \varepsilon_{b_1 b_2}{}^{e_1 .. c_{24}} G_{e_1,e_2 .. c_{24},a}] 
\dot =  {11 \over 12} e_3  \varepsilon_{b_1 b_2}{}^{e_1 .. c_{24}} D_{e_1 e_2 .. e_{23}} \Lambda_{e_{24}a} 
$$
$$
- {23 \over 24} e_3  \varepsilon_{b_1 b_2}{}^{e_1 .. e_{24}}  D_{a e_1 .. e_{22}} \Lambda_{e_{23} e_{24}} + {11 \over 12} e_3  \varepsilon_{b_1 b_2}{}^{e_1 .. e_{24}} \Lambda_{a e_1 .. e_{21}} \overline{D}_{e_{22} e_{23} e_{24}}  
$$
$$
- {11 \over 8} e_3  \varepsilon_{b_1 b_2}{}^{e_1 .. e_{24}} \Lambda_{e_1 .. e_{22}} \overline{D}_{e_{23} e_{24} a}  + 6 \cdot 11 \cdot 23!  e_3  e_2 G_{[b_1,b_2 e_1 .. e_{21}]} \Lambda^{e_1 .. e_{21}}{}_a $$
$$
 - {11 \cdot 23! \over 2} e_3  e_2 \eta_{a[b_1} \Lambda^{e_1 .. e_{22}} G_{[b_2],e_1 .. e_{22}]} + {e_3  \over e_2} \Lambda^e{}_a \overline{G}_{[e,b_1 b_2]} - {1 \over 12} {e_3 \over e_2} \eta_{a[b_1} \Lambda^{e_1 e_2} \overline{G}_{[b_2],e_1 e_2]} 
.  \eqno(4.2.5)$$
\par 
The variation of the gravity-dual gravity relation  (4.2.4) under $I_c(K_{27})$  is given by the sum of (4.2.3) and (4.2.5) and we find that 
$$
\delta {\cal D}_{a,b_1 b_2} \dot =  {11 \over 12} e_3 \varepsilon_{b_1 b_2}{}^{e_1 .. e_{24}} D_{e_1 e_2 .. e_{23}} \Lambda_{e_{24}a} - {23 \over 24} e_3 \varepsilon_{b_1 b_2}{}^{e_1 .. c_{24}}  D_{a e_1 .. c_{22}} \Lambda_{c_{23} c_{24}}  $$
$$
+ {11 \over 12} e_3 \varepsilon_{b_1 b_2}{}^{e_1 .. c_{24}} \Lambda_{a e_1 .. e_{21}} \overline{D}_{e_{22},e_{23} e_{24}} - {11 \over 8} e_3 \varepsilon_{b_1 b_2}{}^{e_1 .. c_{24}} \Lambda_{e_1 .. c_{22}} \overline{D}_{c_{23} c_{24} a} $$
$$
+ (6 + {e_3 \over e_2}) \Lambda^e{}_a \overline{G}_{[b_1,b_2 e]} - {1 \over 12} (6 + {e_3 \over e_2}) \eta_{a[b_1} \overline{G}_{[b_2],a_1 a_2]} \Lambda^{e_1 e_2}$$
$$
+ 11 \cdot 23! (1 + 6 e_3 e_2) [ \Lambda^{e_1 .. e_{21}}{}_a G_{[e_1,e_2 .. e_{22} b_1 b_2]}  - {1 \over 12}  \eta_{a[b_1} G_{[b_2],e_1 .. e_{22}]} \Lambda^{e_1 .. e_{22}}] $$
$$
 - 6 \Lambda^e{}_{[b_1} \tilde{\tilde{D}}_{[b_2],ae]}  + 11 \cdot 23! \Lambda^{e_1 .. e_{21}}{}_{[b_1} \tilde{\tilde{D}}_{[b_2],e_1 .. e_{21}a]}  
  \eqno(4.2.6)$$
In the first two lines we see the desired duality relations. The third and fourth lines vanish on taking the values for $e_1,e_2,e_3$ in equations (4.0.1-4.0.3).  Indeed they only vanish for these values, so demonstrating the impressive consistency of the theory. The fifth line involves duality relations between the Kalb-Ramond fields at levels one and three. These are of the generic form 
$$
\tilde{\tilde{D}}_{b_1 b_2 b_3} \dot =  \overline{G}_{[b_1,b_2 b_3]} + e_4 \varepsilon_{[b_1
}{}^{e_1 .. e_{25}} G_{e_1,e_2 \ldots e_{25},|b_2 b_3]} \ \ , 
\eqno(4.2.7)$$
$$
\tilde{\tilde{D}}_{b_1 b_2 .. b_{23}} \dot = G_{[b_1,b_2 .. b_{23}]} + e_5 \varepsilon_{[b_1|}{}^{e_1 .. e_{25}} G_{e_1,e_2 \ldots  e_{25},|b_2 .. b_{23}]}  \ \ , 
\eqno(4.2.8)
$$ 
The coefficients $e_4$ and $e_5$ are determined by the symmetries of the non-linear realisation but here we do not need to know what they are, just that the duality relations of this type exist.  We regard  equation (4.2.6) as an equivalence relation meaning that it holds up to terms that can be taken to be local Lorentz transformations. 
\par
When varying the duality relation of equation (4.2.4) we did not include the level three variations of the dual graviton which were indicated by $+\ldots$ in equation (3.20). The parameter $\Lambda^{a_1 a_2}$ has levels $(0,\pm 1) $ and so  in the variation of the dual graviton we will find a term of level $(1,2)$. Looking at equation (2.2) we see that it should  contain the Cartan form $G_{a, b_1\ldots b_{24}, c_1c_2}$. Indeed if this Cartan from has a  variation of the form 
$$
\delta  G_{[b_1, \ldots b_{24}], a}\propto \Lambda ^{f_1f_2} G_{[b_1\ldots b_{24}] f_1,a f_2}
\eqno(4.2.9)$$
then, using the duality relation (4.2.7), we can indeed cancel the second to last term in equation (4.2.6). In a similar way one can cancel the  last term in equation (4.2.6) 
\par 
On taking the values of $e_1,e_2,e_3$ in equations (4.0.1-4.0.3) we  find that 
$$
\eqalign{
\delta {\cal D}_{a,b_1 b_2} &\dot = - {11 \over 12} \varepsilon_{b_1 b_2}{}^{e_1 .. c_{24}} D_{e_1 e_2 .. e_{23}} \Lambda_{e_{24}a} + {23 \over 24}  \varepsilon_{b_1 b_2}{}^{e_1 .. c_{24}}  D_{a e_1 .. e_{22}} \Lambda_{e_{23} e_{24}} \cr
&- {11 \over 12} \varepsilon_{b_1 b_2}{}^{e_1 .. e_{24}} \Lambda_{a e_1 .. e_{21}} \overline{D}_{e_{22} e_{23} e_{24}} + {11 \over 8}  \varepsilon_{b_1 b_2}{}^{e_1 .. e_{24}} \Lambda_{e_1 .. e_{22}} \overline{D}_{e_{23} e_{24} a} \cr
& - 6 \Lambda^e{}_{[b_1} \tilde{\tilde{D}}_{b_2]ae}  + 11 \cdot 23! \Lambda^{e_1 .. e_{21}}{}_{[b_1} \tilde{\tilde{D}}_{b_2]e_1 .. e_{21}a} 
} \eqno(4.2.10)$$
Thus the gravity-dual gravity relation (4.2.4) varies into other duality relations and so we can set all the duality relations to zero.


\medskip
{{\bf 4.3 Dilaton duality}}
\medskip
We now consider a duality relation involving the dilaton field $\phi$ at level zero, and the lowest level field that can possibly by considered dual to the dilaton. This is the  field $A_{a_1 .. a_{24}}$ at level two and so we consider such a  duality relation which, on grounds of Lorentz invariance, must be of the form 
$$
D_a \equiv  G_a + e_1 \varepsilon_a{}^{b_1 .. b_{25}} G_{b_1,b_2 .. b_{25}} = 0
 \eqno(4.3.1)$$
\par
As in the previous sections we must add  $l_1$ terms containing  derivatives with respect to the higher level coordinates to the Cartan form $G_a $ such that its transformations under the $I_c(K_{27})$ of equation (3.20) involves curls of the usual spacetime indices. 
 The  $l_1$ extension of $G_a$ is given by 
$$
{\cal G}_a \equiv  G_a + {1 \over 15} \hat{\overline{G}}^{e,}{}_{ea} - {22 \over 3}  \hat{G}^{e_2 .. e_{22},}{}_{e_2 .. e_{22}a} 
\eqno(4.3.2)$$
and its  variation under the $I_c(K_{27})$  is given by 
$$
\delta {\cal G}_a =  \Lambda^{e_1 e_2} \overline{G}_{[a,e_1 e_2]} - {23! \over  6} \Lambda^{e_1 .. e_{22}} G_{[a,e_1 .. e_{22}]}  
 \eqno(4.3.3) $$
 The variation of the second term of equation (4.3.1) does not need an $l_1$ extension due to the presence of the epsilon symbol. 
We take the extended scalar duality relation to be given by 
$$
\eqalign{
{\cal D}_a \equiv {\cal G}_a + e_1 \varepsilon_a{}^{b_1 .. b_{25}} G_{b_1,b_2 .. b_{25}} 
} \eqno(4.3.4) $$
\par
The variation of the second term in the duality relations  of equation (4.3.4) is given by equation (3.20) and this can be rewritten as follows 
$$
\eqalign{
\delta [e_1 \varepsilon_a{}^{b_1 .. b_{25}} G_{[b_1,b_2 .. b_{25}]}] &= + {e_1 \over e_2} \Lambda^{e_1 e_2}  \overline{D}_{[a e_1 e_2]} - e_1 \varepsilon_a{}^{b_1 .. b_{25}} \Lambda_{b_1 .. b_{22}}  \overline{D}_{[b_{23} b_{24} b_{25}]} \cr
& \ \ \  - {e_1 \over e_2} \Lambda^{e_1 e_2}  \overline{G}_{[a,e_1 e_2]} + 3! 23! e_1 e_2 \Lambda^{e_1 .. e_{22}}  G_{[a,e_1 .. e_{22}]} } 
\eqno(4.3.5) $$
Using this result we find that the variation of the scalar duality relation can be written as 
$$
\eqalign{
\delta {\cal D}_a &=  {e_1 \over e_2} \Lambda^{e_1 e_2}  \overline{D}_{a e_1 e_2} - e_1 \varepsilon_a{}^{b_1 .. b_{25}} \Lambda_{b_1 .. b_{22}}  \overline{D}_{ b_{23} b_{24} b_{25}} \cr
& \ \ \ +  (1 - {e_1 \over e_2}) \Lambda^{e_1 e_2}  \overline{G}_{[a,e_1 e_2]} + 3! 23! (e_1 e_2  - {1 \over 3! \cdot 3!} ) \Lambda^{e_1 .. e_{22}} G_{[a,e_1 .. e_{22}]}}
 \eqno(4.3.6) $$
If we use the values for $e_1,e_2,e_3$ of equations (4.0.1-4.0.3) this becomes 
$$
\delta {\cal D}_a =  {1 \over 2} \Lambda^{e_1 e_2}  \overline{D}_{[a,e_1 e_2]} - {1 \over 12} \varepsilon_a{}^{b_1 .. b_{25}} \Lambda_{b_1 .. b_{22}}  \overline{D}_{[ b_{23},b_{24} b_{25}]}  
 + {1 \over 2} \Lambda^{e_1 e_2}  \tilde{\tilde{D}}_{a,e_1 e_2} - {23! \over 12}  \Lambda^{e_1 .. e_{22}} \tilde{\tilde{D}}_{a,e_1 .. e_{22}}
\eqno(4.3.7) $$
In the first line we find the duality relations of equations (4.0.1-4.0.3).  While in the second line we have used  the duality relations of 
equations (4.2.7) and (4.2.8) following a similar argument to their use at the end of  the previous section. Thus the scalar duality relation varies into the other duality relations. 


\medskip
{{\bf 5. Second order equations of motion}}
\medskip
In this section we will derive the equations of motion for the spin zero field $\phi$, the dilaton, and the two form $A_{a_1a_2}$ in two ways. One way is to  use  the duality relations derived in section four and apply another derivative so as to get rid of the dual field to find equation of motion which is second order in derivatives. The second way is to start from scratch and derive the equations of motion using the symmetries of the non-linear realisation. In the variations of the two form equation we will find the equation of motion of the graviton. 
\medskip
{{\bf 5.1 The dilaton equation of motion}}
\medskip
\medskip
{{\bf 5.1.1 Derivation from scalar duality relation}}
\medskip
The scalar duality relation of equation (4.0.1) in space-time indices reads as
$$
D^{\mu} = G^{\mu} + e_1 (\det e)^{-1} \varepsilon^{\mu \nu_1 .. \nu_{25}} G_{\nu_1,\nu_2 .. \nu_{25}} 
\eqno(5.1.1.1)$$
 The Cartan form of $G_{\nu_1,\nu_2 .. \nu_{25}}$ is given in equation (3.14) and when written in terms of world indices it reads. 
$$
G_{\nu_1,\nu_2 .. \nu_{25}} = (\det e)^{{1 \over 2}} (\partial_{\nu_1} A_{\nu_2 .. \nu_{25}} - A_{[\nu_2 \nu_3} \partial_{|\nu_1|} A_{\nu_4 .. \nu_{25}]}) 
\eqno(5.1.1.2)$$
The $(\det e)^{{1 \over 2}}$ factor in front arises from the inverse vielbein $(E^{-1})_A{}^{\Pi}$ as defined below equation (3.11). Taking the derivative  $\partial_{\mu}[(\det e)^{{1 \over 2}} D^{\mu}]$ we find  that the field $ A_{\nu_1\ldots \nu_{24}} $ drops out to leave the following second order equation for the scalar field $\phi$:
$$
E \equiv \partial_{\mu}[(\det e)^{{1 \over 2}} G^{\mu}] - e_1 (\det e)^{-1} \varepsilon^{\nu_1 .. \nu_{26}} G_{\nu_1,\nu_2 \nu_3} G_{\nu_4,\nu_5 .. \nu_{26}}=0
 \eqno(5.1.1.3)$$

\medskip
{{\bf 5.1.2 Derivation of scalar equation  from symmetry}}
\medskip
To derive the equations of motion from the symmetries of the non-linear realisation  we follow the now well trodden path using  the Cartan forms which only transform under $I_c(K_{27})$, see references  [8,9] for a discussion. 
We will take as our starting point the scalar equation derived in the last section  and take its variation under the symmetries of the non-linear realisation. In this section we take the  coefficient $e_1$   to be arbitrary.  We will  not take $E$ to vanish,  but rather demand that it be one member of a set of quantities   which transform into each other. As a result we can set all these quantities to zero. Should our starting equation (5.1.3.1)  not belong to this set of quantities it would have to be discarded.  
\par
 In tangent space indices equation (5.1.3.1)  becomes 
$$
\eqalign{
E &\equiv   \{ (\det e)^{{1 \over 2}} e_a{}^{\mu} \partial_{\mu} G^a - G_{c,a}{}^c G^a + {1 \over 2} G_{a,c}{}^c G^a \} - e_1 \varepsilon^{e_1 .. e_{26}} G_{e_1,e_2 e_3} G_{e_4, e_5 .. e_{26}} \cr
&= E_1 + E_2 =0}
\eqno(5.1.2.1)$$
where $E_1$ is the term in curly brackets.
The variation  of $\delta E_1$ can be simplified by noting that it is can be written as 
$$
\delta E_1 = \partial_{\mu}[(\det e)^{{1 \over 2}} (\delta G^{a}) e_a{}^\mu] - (\delta G_{c,a}{}^c) G^a + {1 \over 2} (\delta G_{a,c}{}^c) G^a   \eqno(5.1.2.2)$$
We find that its variation under $I_c(K_{27})$ transformations of equation (3.20 ) is given by 
$$
\delta E_1 = \partial_\mu\{ (\det e)^{{1 \over 2}} ({1\over 3} \Lambda^{\tau_1\tau_2} \bar G^\mu{}_{,\tau_1\tau_2 }
- {22!\over 6} \Lambda^{\tau_1\ldots \tau_{22}} G^\mu{}_{,\tau_1\ldots \tau_{22} })\}
$$
$$
-4\Lambda ^{ec} \bar G_{c,eb}-22! \cdot 21 \Lambda ^{e_1\ldots e_{21} c} G_{c,e_1\ldots e_{21} b}
\eqno(5.1.2.3)$$
We may write the very first term as 
$$
 \partial_\mu\{ (\det e)^{{1 \over 2}} ( \bar G^{[\mu ,\tau_1\tau_2 ]}  \Lambda_{\tau_1\tau_2}-{2\over 3} \bar G^{\tau_1,\tau_2 \mu  }\Lambda_{\tau_1\tau_2}) \}
\eqno(5.1.2.4)$$
Using equation (3.25) the last term can be cancelled by adding  to $E_1$ the $l_1$  term 
$$
 +{1\over  15 } \partial_\mu\{ ( \det e)^{{1 \over 2}} \hat {\bar G}^e{}_{,e}{}^{\mu}\}
\eqno(5.1.2.5)$$
while the first  term can be written as 
$$
\partial_\mu\{ (\det e)^{{1 \over 2}}  \bar G^{[\mu,\tau_1\tau_2] }\} \Lambda^{\tau_1\tau_2}\}
+(\det e)^{{1 \over 2}}G_\mu \bar G^{[\mu,\tau_1\tau_2]} \Lambda^{\tau_1\tau_2}
$$
$$
+4(\det e)^{{1 \over 2}} \bar G^{[\mu, \tau_1\tau_2]} G _{\mu, (\tau_1\lambda)}  \Lambda^{\lambda}{}_{ \tau_2}
\eqno(5.1.2.6)$$
In deriving this equation we have used equation (3.18) and (3.19)  which can be written as 
$$
\partial_{\mu} \Lambda^{\nu_1 \nu_2} = G_{\mu} \Lambda^{\nu_1 \nu_2} \ \ \ , \ \ \ 
\partial_{\mu} \Lambda^{\nu_1 .. \nu_{22}} = - G_{\mu} \Lambda^{\nu_1 .. \nu_{22}} \ \ \ . 
\eqno(5.1.2.7)$$
The last terms in equations (5.1.2.6) can be rewritten using the expression for the spin connection of equation (4.04) as 
$$
-2\det e \bar G^{[\mu, \tau_1\tau_2]}  \omega_{_\lambda,(\mu, \tau_1)}  \Lambda^{\lambda}{}_{ \tau_2}
\eqno(5.1.2.8)$$
plus a term that can be removed by adding an $l_1$ term to $E_1$. 
\par
Proceeding in a similar way for the other terms in $E_1$ we find that its variation can be written as 
$$
\delta {\cal E}_1 = \Lambda_{\nu_1 \nu_2} E^{\nu_1 \nu_2}- {23! \over 6} \Lambda_{\nu_1 \ldots \nu_{22}} \partial_{\mu}[(\det e)^{{1 \over 2}} D^{\mu\nu_1 .. \nu_{22}}] 
+ 2\Lambda_{\nu_1 \nu_2} G_{\mu} \overline{G}^{[\mu,\nu_1 \nu_2]} 
 $$
 $$
   -2 G^{[e, e_2e_3]} \omega _{d, e_1e_2}\Lambda^{d}{}_{e_3}
+ {1\over 3\cdot 6 e_2}  \varepsilon^{b_1 .. b_{26}} \overline{G}_{b_1,b_2 b_3} G_{b_4} \Lambda_{b_5 .. b_{26}}  \eqno(5.1.2.9)
$$
$$
+{11\over 18 e_2}  \varepsilon^{b_1 .. b_{26}} \overline{G}_{b_1,b_2 b_3} \Lambda^e{}_{b_4 .. b_{24}} (\det e)^{{1 \over 2}} \omega_{e,b_{25} b_{26}
} $$
We define the quantity 
$$
E^{\nu_1 \nu_2} \equiv \partial_{\mu}[ (\det e)^{{1 \over 2}} \overline{G}^{[\mu,\nu_1 \nu_2]}] - G_{\mu} \overline{G}^{[\mu,\nu_1 \nu_2]}  \eqno(5.1.2.10)$$
which is  the  second order equation of motion for the Kalb-Ramond field $A_{\nu_1 \nu_2}$ when it is taken  to vanish. 

The $l_1$-extension of $E_1$ is given by 
$$
{\cal E}_1=E_1+{1 \over 15} \partial_{\mu} \{ (\det e)^{{1 \over 2}} \hat{\overline{G}}^{\tau_2,}{}_{\tau_2}{}^{\mu} \} - {1 \over 5} \overline{G}^{[\mu,\tau_1 \tau_2]} \hat{G}_{\tau_2,(\mu \tau_1)}
- {2 \over 5} \hat{\overline{G}}^{c,}{}_{, cb} G^b  
$$
$$
-  {22 \over 3} \partial_{\mu} \{ \hat G^{\tau_1,\tau_2 .. \tau_{22} \mu} \Lambda_{\tau_1 .. \tau_{22}} (\det e)^{{1 \over 2}} \}
- { 21 \over 11}  \hat G^{e_2 .. e_{22},}{}_{e_2 .. e_{22}b} G^b \eqno(5.1.2.11)
$$
$$
- {2 \over 21!} \varepsilon^{a_1 .. a_{26}} \overline{G}_{[a_1,a_2 a_3]} \hat{G}_{a_5 .. a_{24},[a_{25} a_{26}]} ({1 \over 6 \cdot 6 e_2} ) 
$$
In deriving equation (5.1.9) we have used the definition of the quantities that appear in the duality relations in section 4.
\par
We now apply the same strategy to the second term in equation (5.1.2.1) to find that 
$$
\delta {\cal E}_2 = - {25! \over 36} e_1 \varepsilon^{b_1 .. b_{26}}  \Lambda^{e_1 .. e_{22}}  D_{e_1,e_2 .. e_{22} b_1 b_2 b_3 } {G}_{b_4,b_5 .. b_{26}} 
$$
$$
+ {22 \cdot 24! \over 3} e_1 \Lambda^{e_1 .. e_{22}} \varepsilon^{a_1 .. a_{26}} D_{a_1 a_2 a_3 e_1 .. e_{21},e_{22}}  {G}_{b_4,b_5 .. b_{26}}
$$
$$
+ 100 e_1 \Lambda^{e_1 e_2} \varepsilon^{a_1 .. a_{26}} \overline{G}_{[a_1,a_2 a_3]} D_{a_4 .. a_{26} e_1 e_2}
+ 48 e_1 \Lambda^{e_1 e_2} \varepsilon^{a_1 .. a_{26}} \overline{G}_{[a_1,a_2 a_3]} D_{a_4 .. a_{26} e_1,e_2}
$$
$$
-G_b \overline{G}^{[b,e_1 e_2]} \Lambda_{e_1 e_2} (\det e)^{{1 \over 2}} (1 + 2 {e_1 \over e_2})
+ \overline{G}^{[a_1,a_2 a_3]} \Lambda^e{}_{a_1} (\det e)^{{1 \over 2}} \omega_{e,a_2 a_3} (2{e_1 \over e_2}  - {2 \cdot 6 e_1 \over e_3})$$
$$
+ 22 \varepsilon^{a_1 .. a_{26}} G_{a_1,a_2 a_3} \Lambda^e{}_{a_4 .. a_{24}} (\det e)^{{1 \over 2}} \omega_{e,a_{25} a_{26}} (-e_1 + {e_1 \over 6 e_2 e_3}) $$
$$
- \varepsilon^{a_1 .. a_{26}} \overline{G}_{[a_1,a_2 a_3]} G_{a_4} \Lambda_{a_5 .. a_{26}} ( 2 e_1 - {1 \over  e_2 36}) \eqno(5.1.2.12)$$
Where the $l_1$ extension of $E_2$ is given by 
$$
{\cal E}_2=
+ {1 \over 10} {e_1 \over e_2} \overline{G}^{[a_1,a_2 a_3]} \hat{G}_{a_1,a_2 a_3}  
+ {4 \over 5} e_1 \varepsilon^{a_1 .. a_{26}} \overline{G}_{[a_1,a_2 a_3]} \hat{G}^{e,}{}_{e a_4 .. a_{26}}
$$
$$
- {2 \over 5} e_1 \varepsilon^{a_1 .. a_{26}} \overline{G}_{[a_1,a_2 a_3]} \hat G^{e,}{}_{a_4 .. a_{26},e}
- {4 \cdot 23\cdot 22 \over 3 } e_1 \varepsilon^{a_1 .. a_{26}} \hat{G}^{e_2 .. e_{22},}{}_{e_2 .. e_{22} a_1 a_2 a_3} G_{a_4,a_5 .. a_{26}}
$$
$$
+ 23  \cdot 22\cdot 14 \ e_1 \varepsilon^{a_1 .. a_{26}} \hat G^{e_2 .. e_{22},}{}_{e_2 .. e_{21} a_1 a_2 a_3,e_{22}}G_{a_4,a_5 .. a_{26}}
$$
$$
- {2 \over \cdot 21!} \varepsilon^{a_1 .. a_{26}} \overline{G}_{[a_1,a_2 a_3]} G_{a_4}\hat{G}_{a_5 .. a_{24},[a_{25} a_{26}]} (  {e_1 \over 6 e_2 e_3} - e_1) 
 \eqno(5.1.2.13)$$
In deriving equation (5.1.2.12) we have used the definitions of the quantities that define the duality relations. 
\par
Finally we may find the variation of the scalar equation of motion of equation (5.1.2.1) under the $I_c(K_{27})$ transformations of equation (3.20) is given by the sum of the expressions in equations (5.1.2.9) and (5.1.2.12) to be 
$$
\delta {\cal E} = \Lambda_{\nu_1 \nu_2} E^{\nu_1 \nu_2} - {23! \over 6} \Lambda_{\nu_1 .. \nu_{22}} \partial_{\mu}[(\det e)^{{1 \over 2}} D^{\mu\nu_1 .. \nu_{22}}] 
$$
$$
- {25! \over 3 \cdot 12} e_1 \varepsilon^{b_1 .. b_{26}}  \Lambda^{e_1 .. e_{22}}  D_{e_1,e_2 .. e_{22} b_1 b_2 b_3 } 
{G}_{b_4,b_5 .. b_{26}}
$$
$$
+ 22 {24! \over 3} e_1 \Lambda^{e_1 .. e_{22}} \varepsilon^{a_1 .. a_{26}} D_{a_1 a_2 a_3 e_1 .. e_{21},e_{22}} {G}_{b_4,b_5 .. b_{26}}
+ 100 \ e_1 \Lambda^{e_1 e_2} \varepsilon^{a_1 .. a_{26}} \overline{G}_{[a_1,a_2 a_3]} D_{a_4 .. a_{26} e_1 e_2}$$
$$
+ 48\  e_1 \Lambda^{e_1 e_2} \varepsilon^{a_1 .. a_{26}} \overline{G}_{[a_1,a_2 a_3]} D_{a_4 .. a_{26} e_1,e_2}+
G_b \overline{G}^{[b,e_1 e_2]} \Lambda_{e_1 e_2} (\det e)^{{1 \over 2}} (1 + 1 - {e_1 \over e_1} - 2 {e_1 \over e_2})$$
$$
+ \overline{G}^{[a_1,a_2 a_3]} \Lambda^e{}_{a_1} (\det e)^{{1 \over 2}} \omega_{e,a_2 a_3} (2{e_1 \over e_2} - 2 - {2 \cdot 6 e_1 \over e_3})
$$
$$
+ 22 \varepsilon^{a_1 .. a_{26}} G_{a_1,a_2 a_3} \Lambda^e{}_{a_4 .. a_{24}} (\det e)^{{1 \over 2}} \omega_{e,a_{25} a_{26}} (-e_1 + {e_1 \over 6 e_2 e_3} + {1 \over 6 \cdot 6 e_2}) $$
$$
+ \varepsilon^{a_1 .. a_{26}} \overline{G}_{[a_1,a_2 a_3]} G_{a_4} \Lambda_{a_5 .. a_{26}} (- 2 e_1 + {1 \over 6 \cdot 3 e_2} - {e_1 \over e_1 e_2 6 \cdot 6})
 \eqno(5.1.2.14)$$
The last four terms must vanish and solving for  $e_1,e_2,e_3$ we find that they take the same values as in  equations (4.0.1) --- (4.0.3). 
The  $l_1$-extension of $E$ of equation (5.1.4) is just  the sum of those found previously: 
$$
{\cal E} = {\cal E}_1 + {\cal E}_2 . 
\eqno(5.1.2.15)$$
\par
The final result for the variation of the scalar equation is then given by 
$$
\delta {\cal E} = \Lambda_{\nu_1 \nu_2} E^{\nu_1 \nu_2} - {23! \over 6} \Lambda_{\nu_1 .. \nu_{22}} \partial_{\mu}[(\det e)^{{1 \over 2}} D^{\mu\nu_1 .. \nu_{22}}] $$
$$
- {25! \over 36} {1\over 12} \varepsilon^{b_1 .. b_{26}}  \Lambda^{e_1 .. e_{22}}  D_{e_1,e_2 .. e_{22} b_1 b_2 b_3 } 
{D}_{b_4,b_5 .. b_{26}} \eqno(5.1.2.16)
$$
$$
+ 22 {24! \over 3}  \Lambda^{e_1 .. e_{22}} \varepsilon^{a_1 .. a_{26}} D_{a_1 a_2 a_3 e_1 .. e_{21},e_{22}} {D}_{b_4,b_5 .. b_{26}} $$
$$
+  {25\over 3}  \Lambda^{e_1 e_2} \varepsilon^{a_1 .. a_{26}} \overline{G}_{[a_1,a_2 a_3]} D_{a_4 .. a_{26} e_1 e_2}
+ 4  \Lambda^{e_1 e_2} \varepsilon^{a_1 .. a_{26}} \overline{G}_{[a_1,a_2 a_3]} D_{a_4 .. a_{26} e_1,e_2}$$
The first term contains $E^{\tau_1\tau_2}$ which can be identified with  the two form equation of motion when set to zero. In all the other terms we find  expressions that occur in the  duality relations which also vanish. 
Thus $E$ as defined in  equation  (5.1.2.1) varies into the equation of motion of the two form and  duality relations and so we can conclude that  the scalar equation of motion is indeed given by equation  (5.1.2.1). 
\par
We may rewrite the scalar equation as 
$$
E = \partial_{\nu}[(\det e)^{{1 \over 2}} G^{\nu}] + {1 \over 2} \overline{G}_{[c_1 ,c_2 c_2]} \overline{G}^{[c_1,c_2 c_3]} 
- {1 \over 12} \varepsilon^{e_1 .. e_{26}} \overline{G}_{[e_1,e_2 e_3]} {D}_{e_4 e_5 .. e_{26}}
$$
$$
=  \partial_{\nu}[(\det e)^{{1 \over 2}} G^{\nu}] + {1 \over 2} \overline{G}_{[c_1 ,c_2 c_2]} \overline{G}^{[c_1,c_2 c_3]} 
 \eqno(5.1.2.17)$$
where in the last line we set the two form duality relation to zero. 

\medskip
{{\bf 5.2 Second order Kalb-Ramond equation of motion}} 
\medskip
{{\bf 5.2.1 Derivation from the  Kalb-Ramond duality relation}}
\medskip
The lowest order duality relation involving the two  field $A_{a_1a_2}$ of equation (4.0.2) is first order in derivatives and when expressed in terms of  
it  reads as
$$
D^{\mu \nu_1 \nu_2} \equiv  \overline{G}^{[\mu,\nu_1 \nu_2]} + e_2 (\det e)^{-1} \varepsilon^{\mu \nu_1 \nu_2 \rho_1 .. \rho_{23}} G_{\rho_1,\rho_2 .. \rho_{23}} =0 
\eqno(5.2.1.1)$$
 The Cartan form of $G_{\rho_1,\rho_2 .. \rho_{23}}$ of  equation (3.14) when written with world volume indices is  given by 
$$
G_{\rho_1,\rho_2 .. \rho_{22}} = e^{\phi} (\det e)^{{1 \over 2}} \partial_{\rho_1} A_{\rho_2 .. \rho_{22}}
 \eqno(5.2.1.2)$$
Taking a derivative, that is, evaluating $\partial_{\mu} [(\det e)^{{1 \over 2}} \overline{D}^{\mu,\nu_1 \nu_2}]$,  the dual field $A_{\mu_1\ldots \mu_{22}}$ drops out and we find the equation of motion for the Kalb-Ramond field:
$$
E^{\nu_1 \nu_2} = \partial_{\mu}[ (\det e)^{{1 \over 2}} \overline{G}^{[\mu,\nu_1 \nu_2]}]
+e_2\epsilon^{\mu_1\mu_2 \nu_1\ldots \nu_{24}} G_{\nu_1}G_{\nu_2, \ldots \nu_{24}}
\eqno(5.2.1.3)$$
Using the duality relations for the two form we may write it as 
$$
E^{\nu_1 \nu_2} =\partial_{\mu}[ (\det e)^{{1 \over 2}} \overline{G}^{[\mu,\nu_1 \nu_2]}] - G_{\mu} \overline{G}^{[\mu,\nu_1 \nu_2]} = 0 
\eqno(5.2.1.4)$$
This agrees with the two form field equation we found by varying the scalar equation in the previous section. 
\medskip
{{\bf 5.2.2 Derivation of  the  Kalb-Ramond using symmetry}}
\medskip
In  section 5.1.2 we found the second order Kalb-Ramond equation of motion in equation (5.1.2.10) which can be written as in equation (5.1.2.1.3) by varying the second order scalar equation of motion $E$ under the local $I_c(K_{27})$ transformations of the non-linear realisation. In tangent space this equation becomes 
$$
E^{a_1 a_2} = \{(\det e)^{{1 \over 2}} e_b{}^\mu  \partial_{\mu}[ \overline{G}^{[b, a_1a_2]}] -2G_{b,c}{}^{[a_1} \bar G ^{[ b,c |a_2]}
$$
$$
- G_{c,b} {}^c\overline{G}^{[b,a_1 a_2]} +{1\over 2} G_{b,c}{}^c \bar{G}^{[b,a_1 a_2]} \}
+e_2\epsilon^{a_1a_2 b_1\ldots b_{24}} G_{b_1}G_{b_2, \ldots b_{24}}= 0 
\eqno(5.2.2.1)$$
\par 
The variation of the first three terms  in the curly brackets can be written as 
$$
e_{\mu_1}{}^{[a_1}e_{\mu_2}{}^{a_2]}  \partial_{\nu}\{ (\det e)^{{1 \over 2}} \delta \overline{G}^{[b, c_1c_2]}  e_b{}^\nu 
e_{c_1}{}^{\mu_1}e_{c_2}{}^{\mu_2}\}-2(\delta G_{b,c}{}^{[a_1} )\bar G ^{[ b,c |a_2]}
$$
$$
- (\delta G_{c,b} {}^c)\overline{G}^{[b,a_1 a_2]} +{1\over 2} (\delta G_{b,c}{}^c) \bar{G}^{[b,a_1 a_2]} 
\eqno(5.2.2.2)$$
Under $I_c(K_{27})$ transformations of equation (3.20) we find that 
$$
\delta {\cal E}^{a_1 a_2} = - {2 \over 3} \Lambda^{a_1 a_2} E - {4 \over 3} \Lambda^{e[a_1} (\det e) \tilde E_e{}^{a_2]} 
$$
$$
+ {25! \over 36} e_{\mu_1}{}^{a_1}e_{ \mu_2}{}^{ a_2} \Lambda_{\rho_1 .. \rho_{22}} \partial_{\nu} \{(\det e)^{{1 \over 2}} D^{\nu \mu_1 \mu_2 \rho_1 .. \rho_{22}} \}
$$
$$
- {23! \over 6} e_2 \varepsilon^{a_1 a_2 b e_1 .. e_{23}} D_{e_1 .. e_{23}} G_{[b,g_1 .. g_{22}]} \Lambda^{g_1 .. g_{22}} \eqno(5.2.2.3)
$$
$$
- 100 e_2 \varepsilon^{a_1 a_2 b e_1 .. e_{23}} \Lambda^{f_1 f_2} D_{e_1 .. e_{23} f_1 f_2} G_b
- 96 e_2 \varepsilon^{a_1 a_2 b e_1 .. e_{23}} \Lambda^{f_1 f_2} D_{e_1 .. e_{23} f_1 , f_2} G_b$$
$$
-  {22\cdot 24!  \over 3} \ e_{\mu_1}{}^{a_1}e_{ \mu_2}{}^{a_2} \partial_{\nu} \{ (\det e)^{{1 \over 2}} D^{\nu \mu_1 \mu_2 e_1 .. e_{21},e_{22}}\Lambda_{e_1\ldots e_{22}} \} $$
$$
+ {11 \over 3} ({1 \over e_3} + {1 \over 6 e_1} - 2 \cdot 3 e_2) G_b \varepsilon^{a_1 a_2 e_1 .. e_{21} f_1 f_2 b} \Lambda^d{}_{e_1 .. e_{21}} (\det e)^{{1 \over 2}} \omega_{d,f_1 f_2}$$
$$
- {2 \over 3} (1 - {e_2 \over 2 e_1}) \Lambda^{a_1 a_2} G^b G_b  - {4 \over 3} ({1 \over e_3} + 1) \Lambda^{c[a_1|} (\det e)^{{1 \over 2}} \omega_{c,}{}^{b|a_2]} \ .$$
where 
$$
(\det e) \tilde{E}_a{}^b \equiv (\det e) R_a{}^b - 9 \overline{G}_{[a,e_1 e_2]} \overline{G}^{[b,e_1 e_2]} + {1 \over 4} \delta_a{}^b \overline{G}_{[e_1,e_2 e_3]} \overline{G}^{[e_1 , e_2 e_3]} - 6 G_a G^b
 \eqno(5.2.2.4)$$
\par 
The $l_1$-extension ${\cal E}^{b_1 b_2}$ of equation (5.2.3.3) is given by 
$$
{\cal E}^{b_1 b_2} ={ E}^{b_1 b_2} 
- {2 \over 5} \hat{\bar G}^{e,}{}_{eb} \bar G^{[b,a_1 a_2]} - 44 \hat{G}^{e_1 .. e_{21},}{}_{e_1 .. e_{21} b} \bar G^{[b,a_1 a_2]}$$
$$
+ {4  \cdot 23 \cdot 22 \over 3}  e_{\mu_1 \mu_2}^{a_1 a_2} \partial_{\nu} \{ (\det e)^{{1 \over 2}}  \hat{G}_{e_2 .. e_{22},}{}^{e_2 .. e_{22} \nu \mu_1 \mu_2} \} $$
$$
- {1 \over 15} e_{\mu_1 \mu_2}^{a_1 a_2}  \hat{\partial}_{\tau} \{ (\det e) \omega_{\tau,}{}^{\mu_1 \mu_2} \}
+ {1 \over 3 \cdot 6 \cdot 21! e_1} \hat{G}^{d_2 .. d_{22},}{}_{[c_1 c_2]}  \varepsilon^{a_1 a_2 c_1 c_2 d_2 .. d_{22} f} G_f
$$
$$
- {1 \over 5} e_{\mu_1 \mu_2}^{a_1 a_2}  \partial_{\nu} \{ (\det e) G^{[\mu_1,}{}^{|\nu|\mu_2]} \}   
+ {2 \over 15} e_{\mu_1 \mu_2}^{a_1 a_2} \hat{\partial}^{\mu_1} \{(\det e)^{{1 \over 2}} G^{\mu_2} \}
$$
$$
+ {2 \over 15}  (\det e) e_{\mu}{}^{[a_1 } (\hat{\partial}^{\mu} \omega_{\nu,}{}^{a_2] b} )e_b{}^{\nu} 
+ {2 \over 5} \hat{\overline{G}}^{[a_1 | }{}_{, bd} \overline{G}^{[b,d|a_2]]} 
+ {1 \over 15} \hat{G}^{e,}{}_{e b} \overline{G}^{[b,a_1 a_2]}$$
$$
+ {22\cdot 11 \over 3}   G^{e_2 .. e_{22},}{}_{e_2 .. e_{22} b} \overline{G}^{[b,a_1 a_2]}
+  22 \cdot 4 \  \hat{G}^{e_2 .. e_{21}[a_1,}{}_{e_2 .. e_{21}bd} \bar G^{[a_2],bd]} $$
$$
+ {1 \over 15} e_2 \varepsilon^{a_1 a_2 b e_1 .. e_{23}} G_{e_1,e_2 .. e_{23}} \hat{G}^{f,}{}_{f b} 
- {4 \over 5} e_2 \varepsilon^{a_1 a_2 b e_1 .. e_{23}} \hat{G}^{f,}{}_{f e_1 e_2 .. e_{23}} G_b $$
$$
- {22 \over 3}  e_2 \varepsilon^{a_1 a_2 b e_1 .. e_{23}} \hat{G}^{g_2 .. g_{22},}{}_{g_2 .. g_{22}b} G_{e_1,e_2 .. e_{23}}
+ {2 \over 21!} e_2 \varepsilon^{a_1 a_2 b e_1 .. e_{23}} \hat{G}_{e_1 .. e_{21},e_{22} e_{23}}$$
$$
- {2 \over 5} {e_2 \over e_3}  \hat{G}^{e}  (\det e)^{{1 \over 2}} \omega_{e,}{}^{a_1 a_2} 
+ {2 \over 5} e_2 \varepsilon^{a_1 a_2 b e_1 .. e_{23}} \hat{G}^{f,}{}_{e_1 .. e_{23},f}$$
$$
+ {1 \over 15} ({e_2\over e_1}-11)\hat{G}^{[a_1} G^{a_2]}  + {1 \over 15}  \hat{G}^c  (\det e)^{{1 \over 2}} \omega_{c,}{}^{a_1 a_2}.
\eqno(5.2.2.5)$$
where $e_{\mu_1 \mu_2}^{a_1 a_2}  = e_{\mu_1}{}^{[ a_1}e_{ \mu_2}{}^{ a_2]}  $. The expressions  involving $\Lambda^{\tau\nu} \partial_\nu$ lead to terms of the form $\Lambda^{\tau\nu} G_{\nu,\alpha}= e^{\tau }{}_d\Lambda^{d c}G_{c, \alpha}$ which can be cancelled by adding ${1\over 10} e^{\tau }{}_d \hat G^d{}_{,\alpha}= {1\over 10} \hat G^\tau{}_{,\alpha}$. 
\par
The last three terms in equation (5.2.2.3) must vanish if the two form equation is to be part of a set of quantities that transform into each other. One finds that the constants $e_1$,  $e_2$, and  $e_3$ must take the values of equations (4.0.1) --- (4.0.3). This yet again demonstrated the very strong internal consistency of the derivation of the equations of motion. Taking these values the 
transformation of the two form equation is given by $$
\delta {\cal E}^{a_1 a_2} = - {2 \over 3} \Lambda^{a_1 a_2} E - {4 \over 3} \Lambda^{e[a_1} (\det e) \tilde E_e{}^{a_2]} 
$$
$$
+ {25! \over 3 \cdot 12} e_{\mu_1}{}^{a_1}e_{ \mu_2}{}^{ a_2} \Lambda_{\rho_1 .. \rho_{22}} \partial_{\nu} \{(\det e)^{{1 \over 2}} D^{\nu \mu_1 \mu_2 \rho_1 .. \rho_{22}} \} 
$$
$$
- {23! \over 36}  \varepsilon^{a_1 a_2 b e_1 .. e_{23}} D_{e_1 .. e_{23}} G_{[b,g_1 .. g_{22}]} \Lambda^{g_1 .. g_{22}} 
- {4 \cdot 25 \over 6} \varepsilon^{a_1 a_2 b e_1 .. e_{23}} \Lambda^{f_1 f_2} D_{e_1 .. e_{23} f_1 f_2} G_b $$
$$
-  16 \varepsilon^{a_1 a_2 b e_1 .. e_{23}} \Lambda^{f_1 f_2} D_{e_1 .. e_{23} f_1 , f_2} G_b
-  {22\cdot 24! \over 3  }e_{\mu_1 \mu_2}^{a_1 a_2} \partial_{\nu} \{ (\det e)^{{1 \over 2}} D^{\nu \mu_1 \mu_2 e_1 .. e_{21},e_{22}}\Lambda_{e_1\ldots e_{22}} \} .\eqno(5.2.6)$$ 
Thus the two form equation transforms into the Einstein equation of motion and duality relations.


\medskip
{{\bf 5.3 Comparison with the low energy effective action of the closed bosonic string}}
\medskip
In section four we found  the equations of motion of the spin zero, the dilaton, the spin two and gravity in equations  (5.1.2.10), (5.1.2.10) and (5.2.2.4) respectively and we collect them here for convience 
$$
E=  \partial_{\nu}[(\det e)^{{1 \over 2}} G^{\nu}] + {1 \over 2} \overline{G}_{[c_1 ,c_2 c_2]} \overline{G}^{[c_1,c_2 c_3]} =0
 \eqno(5.3.1)$$
$$
E^{\nu_1 \nu_2} \equiv \partial_{\mu}[ (\det e)^{{1 \over 2}} \overline{G}^{[\mu,\nu_1 \nu_2]}] - G_{\mu} \overline{G}^{[\mu,\nu_1 \nu_2]}  =0 
\eqno(5.3.2)$$
$$
(\det e) \tilde{E}_a{}^b \equiv (\det e) R_a{}^b - 9 \overline{G}_{[a,e_1 e_2]} \overline{G}^{[b,e_1 e_2]} + {1 \over 4} \delta_a{}^b \overline{G}_{[e_1,e_2 e_3]} \overline{G}^{[e_1 , e_2 e_3]} - 6 G_a G^b=0 
 \eqno(5.3.3)$$
\par
Using the expressions for the Cartan forms of equation (3.14) we find that these equations of motion are given by 
$$
E=\partial_{\mu} ( (\det e) g^{\mu\nu} \partial_\nu \phi) +{1\over 2} e^{-2\phi} F_{\rho_1\rho_2\rho_3}F^{\rho_1\rho_2\rho_3}=0
 \eqno(5.3.4)$$
$$
E^{\nu_1 \nu_2} \equiv \partial_{\mu}[(\det e)e^{-2\phi} {F}^{\mu \nu_1 \nu_2} ] = 0 
\eqno(5.3.5)$$
and 
$$
R_{\mu\nu}- { e^{-2\phi}\over 9} F_{\mu \tau_1\tau_2} F_{\nu}{}^{ \tau_1\tau_2} + {e^{-2\phi} \over 4.9} g_{\mu\nu} F_{ \tau_1\tau_2\tau_3} F^{ \tau_1\tau_2\tau_3}  - 6 \partial_\mu \phi \partial_\nu \phi   =0
\eqno(5.3.6)$$
where $F_{\mu_1\mu_2\mu_3}=3 e^{\phi} G_{[\mu_1,\mu_2\mu_3]}= 3 \partial_{[\mu_1} A_{\mu_2\mu_3]}$. 
\par
Carrying out the changes 
$$
\phi\to {\phi\over 6},\ F_{\mu_1\mu_2\mu_3}\to 3 F_{\mu_1\mu_2\mu_3},\ e_\mu {}^a\to e_\mu {}^a
\eqno(5.3.7)$$
We find that the equations (5.3.4-6)  come from the action 
$$
S = \int d^{26} x \det e \{ R - {1 \over 6} ( \partial^{\mu} \phi) (\partial_{\mu} \phi) - {1 \over 3} e^{-{2 \over 6} \phi} F_{\mu \nu \rho} F^{\mu \nu \rho} \}
 \eqno(5.3.8)$$
This is the well known action for low energy effects of the closed bosonic string. Thus we have shown that the  low energy effective action of the closed bosonic string in 26 dimensions is contained in the non-linear realisation $K_{27} \otimes_s l_1$ with local subalgebra $ I_c(K_{27})$. In particular it emerges if one keeps only the twenty six  coordinates and discards all the other coordinates of the enlarged spacetime. 


\medskip
{\bf 6. Discusion}
\medskip
We have calculated the dynamics that follows from  the non-linear realisation  of the semi-direct product of $K_{27}$ with its vector representation and shown that if we restrict the spacetime to be the usual twenty six dimensions then this is precisely the   low  energy effective action of the closed bosonic string.  The type II superstrings have low energy actions that are uniquely determined by supersymmetry and this ensures that they contain all perturbative and non-perturbative string effects at low energy. These actions are also uniquely  determined by $E_{11}$ symmetry and so one could regard this as leading to the same conclusion. 
Obviously the closed bosonic string does not possess any supersymmetry but its low energy effective action  is determined by the $K_{27}$ symmetry and one could regard  this symmetry as ensuring that it contains all perturbative and non-perturbative effects. 
\par
The branes in M theory are contained in the vector representation of $E_{11}$ [21]. These include branes whose charges  are not found in the supersymmetry algebra but are known to be present. Similarly the branes in the bosonic theory are contained in the vector representation of $K_{27}$. Looking at equation (2.11) we find a point particle ($P_a$), a1 brane ($Z^a$), a 21 brane  ($Z^{a_1 .. a_{21}}$), two 23 branes  ($Z_{\{1\}}^{a_1 \ldots  a_{23}}$ , $Z_{\{2\}}^{a_1 \ldots a_{23}}$) and the Taub Nut brane 
($Z^{a_1 .. a_{22},b}$) as well as higher level branes. Clearly the 1 brane is the elementary string and the 21 brane its dual analogue.  We note that in the vector representation of $K_{27}$ there is a 25 brane at level three. It would be interesting to find what are the properties of all  these  branes.
\par
The non-linear realisations of $E_{11}$  and $K_{27}$ lead to the low energy effective actions of M theory and  the closed bosonic string respectively. A natural question to ask is what underlying theories possess these very large Kac-Moody symmetries. We now comment on this idea in the context of the results of this paper. It is well known that the closed bosonic string, dimensional reduced on tori,  is  invariant under  discrete $D_{26}=O(26,26)$ T duality transformations. One can also formulate the first quantised string action on the world sheet so that it has a $D_{26}$ symmetry by introducing a coordinate $y_\mu$ in addition to the usual coordinate $x^\mu$ [22,23]. This theory can also be derived from an  $E_{11}$ non-linear realisation applicable to branes [24]. 
Thus the closed bosonic string does have  a $D_{26}$ symmetry. 
\par
In the non-linear realisation studied in this paper the $D_{26}$ symmetry in contained in $K_{27}$ as can be seen by deleting node 26 in the $K_{27}$ Dynkin diagram of equation (2.1). Thus the non-linear realisation naturally contains the $D_{26}$ symmetry. The generators of $D_{26}$ are those of the gravity line, nodes 1 to 25 and the generators $R^{a_1a_2}$  and $R_{a_1a_2}$ associated with node 27. Looking at the commutators of equation (2.13) we see that they rotate $P_a$ and $Z^a$ into each other, indeed they belong to the 26 dimensional vector representation of $D_{26}$. As a result their corresponding symmetry transformations   are just a symmetry of the  string. 
\par
The only node in the $K_{27}$ Dynkin diagram not in $D_{26}$ is node 27. This is  associated with the generators $R^{a_1\ldots a_{22}}$ and $R_{a_1\ldots a_{22}}$. Looking at equation (2.13) we see that these generators transform $P_a$ and $Z^a$ into the generators $Z^{a_1\ldots a_{23}}$ and $Z^{a_1\ldots a_{22}, b}$ and taking further commutators  one finds all the generators in the  vector representation. Thus the transformations  corresponding to node 27 will transform the string into all the higher branes. While $D_{26}$ is a symmetry of string theory this makes it clear that $K_{27}$ must be a symmetry of a theory that includes strings and all the branes. As such the underlying theory that has $K_{27}$ as a symmetry must contain string and branes. Such a theory is needed in any case as the branes arise as solitons of the low energy effective action and, when quantised,  they lead to additional degrees of freedom which should be included. 
\par
One could also construct the non-linear realisations of $D_{D-2}^{+++}$ which would lead to a theory in $D$ dimensions. At low levels this theory would contain a graviton, a dilaton and a two form. If one took $D=10$ then one would cover Siegel theory [25] which describes the low energy effective action of the NS-NS sector of the type II superstring but living in a spacetime with coordinates $x^\mu$ and $y_\mu$. This theory is the level zero sector of the non-linear realisation of the non-linear realisation $E_{11}\otimes_s l_1$ [26]. The low energy effective action of the R-R sector of the superstring is just the level one sector of this theory [27]. It is straight forward to find the generators of  $D_{8}^{+++}$ in $E_{11}$ and it is very likely that all of $D_{8}^{+++}$ is contained in $E_{11}$. 
\par
The tachyon could well play an important role in bosonic string and it would be interesting to extend the results of this paper to include the tachyon.

\medskip
{\bf {Acknowledgements}}
\medskip
Peter West wishes to thank Paul Cook and Arkardy Tseytlin for discussions, and also the SFTC for support from Consolidated grant Pathways between Fundamental Physics and Phenomenology, ST/T000759/1. Keith Glennon would like to thank the Dublin Institute for Advanced Studies, and King's College London, for support while portions of this work was carried out as a visitor at each institution, and the Okinawa Institute of Science and Technology for support while portions of this work was carried out as a Postdoctoral Scholar. 
\medskip


\medskip
{{\bf Appendix A: Chevalley-Serre relations of $K_{27}$}}
\medskip
In this appendix we will identify the Serre generators   amongst the generators of the Kac-Moody algebra $K_{27}$. We closely follow pages 532-6 of reference [17] where the same analysis was given for the $E_{11}$ algebra. The Serre generators $E_a$, $F_a$ and $H_a$ of  any Kac-Moody algebra obey the Serre relations 
$$
[H_m,E_n] = A_{mn} E_n \ , \ [H_m,F_n] = - A_{mn} F_n \ , \ [E_m,F_n] = \delta_{mn} H_m \ , \ m, n = 1,\ldots ,27 \ ,
 \eqno(A.1)$$ 
 The Cartan matrix $A_{mn}$  of $K_{27}$ is the $27 \times 27$ matrix 
$$
A = \pmatrix{
2 & -1 & 0 & 0 & 0 & \ldots & 0 & 0 & 0 & 0 & 0 \cr
-1 & 2 & -1 & 0 & 0 & \ldots & 0  & 0 & 0 & 0 & 0 \cr
0 & -1 & 2 & -1 & 0 & \ldots & 0 & 0 & 0 & 0 & 0 \cr
0 & 0 & -1 & 2 & -1 & \ldots & 0 & 0 & 0 & -1 & 0 \cr
0 & 0 & 0 & -1 & 2 & \ldots & 0 & 0 & 0 & 0 & 0 \cr
\vdots & \vdots & \vdots & \vdots & \vdots & \ddots & \vdots & \vdots & \vdots & \vdots & \vdots \cr
0 & 0 & 0 & 0 & 0 & \ldots & 2 & -1 & 0 & 0 & 0 \cr
0 & 0 & 0 & 0 & 0 & \ldots & - 1 & 2 & -1 & 0 & -1 \cr
0 & 0 & 0 & 0 & 0 & \ldots & 0 & -1 & 2 & 0 & 0 \cr
0 & 0 & 0 & -1 & 0 & \ldots & 0 & 0 & 0 & 2 & 0 \cr
0 & 0 & 0 & 0 & 0 & \ldots & 0 & -1 & 0 & 0 & 2 \cr
} \eqno(A.2)
$$
The form of this matrix immediately follows from the $K_{27}$ Dynkin diagram in equation (2.1). 
\par 
In section two  the $K_{27}$ algebra was analysed  by  decomposing it terms of representations of $A_{25}$. The generators found in this way can be classified by two levels,  or a simpler level which is the sum of these two. The  generators of  $K_{27}$ at levels zero, one, and minus one, as listed in equations (2.3) and (2.4), are given by
$$
K^a{}_b \ , R \ ; \ R^{a_1 a_2} \ , R^{a_1 .. a_{22}} \ ; \ R_{a_1 a_2} \ , \ R_{a_1 .. a_{22}} \ ; \quad a_1,a_2,\ldots =1,\ldots , 26
\eqno(A.3)$$
The level zero generators  $K^a{}_b$ generates $GL(26)$. The  commutators of the  $GL{(26)}$ generators with the other   generators in $K_{27}$ can be chosen to be such that they transform as representations of $GL{(26)}$,  as such 
$$
[K^a{}_b,K^c{}_d] = \delta^c{}_b K^a{}_d - \delta^a{}_d K^c{}_b \ , \ [K^a{}_b,R^{cd}] = 2 \delta^{[c}{}_b R^{|a|d]} \ , \ [K^a{}_b,R_{cd}] = - 2 \delta^a{}_{[c} R_{|b|d]} \ , $$
$$
[K^a{}_b,R^{c_1 .. c_{22}}] = 22 \delta^{[c_1}{}_b R^{|a|c_2 .. c_{22}]} \ , \ [K^a{}_b,R_{c_1 .. c_{22}}] = - 22 \delta^a{}_{[c_1} R_{|b|c_2 .. c_{22}]} \ , 
\eqno(A.4)$$
The  $K^a{}_b$ generate   ${\rm GL}(26)$ rather than $A_{25} =SL(26)$ as they  included the generator  $D = \sum_a K^a{}_a$. The generator $R$ has level zero and must  be chosen to commute with the $K^a{}_b$ generators in order that we have twenty seven  commuting generators at level zero which form the Cartan subalgebra. 
\par
To construct the $K_{27}$ algebra we will need to identify the Serre generators in terms of the above low level generators of equation (A.3). The level is preserved by the commutators and it then  follows that all the generators of $K_{27}$ can be found by taking  the multiple commutators of the generators at level zero, one and minus one. Thus the Serre generators must be contained in  the level zero, level one, and level minus one generators of $K_{27}$. The Chevalley-Serre generators of $A_{25}=SL(26)$ must be constructed from the level zero generators and are given by
$$
H_i = K^i{}_i - K^{i+1}{}_{i+1} \ , \ E_i = K^i{}_{i+1} \ , \ F_i = K^{i+1}{}_i \ . \quad i,j=1,\ldots , 25
\eqno(A.5)$$
These are indeed the well known expressions for the $A_{25}$ Serre generators and it is straight forward to show that they satisfy the Serre relations for $A_{25}$. 
\par 
Following the same arguments as given in reference [17],  for the case of $E_{11}$,  we identify
$$
E_{27} = R^{25 \ 26} \ , \ F_{27} = R_{25 \ 26} \ ; \ E_{26} = R^{5 .. 26} \ , \ F_{26} = R_{5 \ .. \ 26} \ ,
 \eqno(A.6)$$
It just remains to find the Cartan subalgebra generators $H_{26}, H_{27}$. These must involve the commuting generators in $K_{27}$ and so can be written as  
$$
H_{26} = K^5{}_5 + .. + K^{26}{}_{26} + \lambda_{26} D + \mu_{26} R  \ , \eqno(A.7)$$
$$
H_{27} = K^{25}{}_{25} + K^{26}{}_{26} + \lambda_{27} D + \mu_{27} R  \ , \eqno(A.8)$$
 The dependence on $K^a{}_b$ is found by requiring that $H_{26}$ and $H_{27}$ obey the Serre relations involving  $H_i, \ i=1, 15$. 
\par
In order to fix the coefficients in these generators, we will need the following commutators between $R$ and the level one and minus-one $K_{27}$ generators of equation (A.6). The commutator of $R$ with any  generator must, as a result of Jacobi identities, preserve the SL(26) nature of that generator. As a result we must have 
$$
\eqalign{
[R,R^{a_1 a_2}] &= + R^{a_1 a_2} \ \ , \ \ [R,R^{a_1 .. a_{22}}] =  - R^{a_1 .. a_{22}} \ \  ,  \cr
[R,R_{a_1 a_2}] &= - R_{a_1 a_2} \ \ , \ \ [R,R_{a_1 .. a_{22}}] =  + R_{a_1 .. a_{22}} \ \ .}
 \eqno(A.9)$$
where in the first relation we can choose  the coefficient  to be 1 by rescaling $R$. The third  relation follows by applying the Cartan Involution to the first relation. 
\par 
The coefficients in equations (A.7) and (A.8) can be fixed by requiring that these Cartan generators obey the correct Serre relations with the generators $E_{26}$ and  $E_{27}$. In particular evaluating the commutators of $H_{26}$ and $H_{27}$ with $E_{26}$ and $E_{27}$  we find that 
$$
20 + 22 \lambda_{26} - \mu_{26} = 0 \ \ , \ \ 2 \lambda_{27} + \mu_{27} = 0 \ \ .
 \eqno(A.10)$$ 
and 
$$
2 + 2 \lambda_{26}  + (20 + 22 \lambda_{26}) = 0 \ \ , \ \ 2 + 22 \lambda_{27} + 2  \lambda_{27} = 0 \ \ \ . 
\eqno(A.11)$$
\par
Solving these four equations for the four unknowns we find that 
$$
\eqalign{
H_{26} &= K^{5}{}_{5} + .. + K^{26}{}_{26} - {11 \over 12} D - {1 \over 6} R \ \ , \cr
H_{27} &= K^{25}{}_{25} + K^{26}{}_{26} - {1 \over 12} D + {1 \over 6} R \ \  .} 
\eqno(A.12)$$
\par
The algebra of $K_{27}$ is constructed at low levels by requiring the generators of equation (2.2) and their negative level equivalents  satisfy the Jacobi identities. However, one also has to require that they satisfy the Serre relations. For example, using equations (A.6) and (A.12), we require that  
$$
[E_{27},F_{27}] = H_{27} = K^{25}{}_{25} + K^{26}{}_{26} - {1 \over 12} D + {1 \over 6} R 
$$
$$
= [R^{25 \ 26},R_{25 \ 26}] = 4 \delta^{[25}{}_{[25} K^{26]}{}_{26]} - {1 \over 6} \delta^{25 \ 26}_{25 \ 26} D + {1 \over 3} \delta^{25 \ 26}_{25 \ 26} R 
\eqno(A.13)$$
As a result we must conclude that 
$$
[R^{a_1 a_2},R_{b_1 b_2}] = 4 \delta^{[a_1}{}_{[b_1} K^{a_2]}{}_{b_2]} - {1 \over 6} \delta^{a_1 a_2}_{b_1 b_2} D + {1 \over 3} \delta^{a_1 a_2}_{b_1 b_2} R 
\eqno(A.14)$$
Similarly using the $[E_{26},F_{26}] = H_{26}$  relation we find that 
$$
[R^{a_1 .. a_{22}},R_{b_1 .. b_{22}}] = 22 \cdot 22! \delta^{[a_1 .. a_{21}}_{[b_1 .. b_{21}} K^{a_{22}]}{}_{b_{22}]} - {11 \over 12} 22! \delta^{a_1 .. a_{22}}_{b_1 .. b_{22}} D - {22! \over 6} \delta^{a_1 .. a_{22}}_{b_1 .. b_{22}} R 
 \eqno(A.15)$$
 To find the $K_{27} $ algebra it suffices to satisfy these Serre relations and then use the Jacobi identities at all higher levels. 


\medskip
{{\bf Appendix B: $K_{27}$ algebra to level two}}
\medskip
In this appendix we give the algebra of all the generators of $K_{27}$ to level two with the exception of the  level two $R^{25,19}$ generator. The difficulties encountered when including this generator are commented on at the end of this sub-appendix.
\par 
The algebra of the non-negative generators to level two is given by
$$
[K^a{}_b,R] = 0 \ , \ [K^a{}_b,R^{c_1 c_2}] = 2 \delta^{[c_1}{}_b R^{|a|c_2]} \ , \ [K^a{}_b,R^{c_1 .. c_{22}}] = 22 \delta^{[c_1}{}_b R^{|a|c_2 .. c_{22}]} \ , $$
$$
 [K^a{}_b,R^{c_1 .. c_{24}}] = 24 \delta^{[c_1}{}_b R^{|a|c_2 .. c_{24}]} \ ,  $$
$$
[K^a{}_b,R^{c_1 .. c_{23},d}] = \delta^{d}{}_b R^{a,c_1 .. c_{23}}  + 23 \delta^{[c_1}{}_b R^{|a|c_2 .. c_{23},d]} \ ,  $$
$$
[R,R] = 0 \ \ , \ \ [R,R^{a_1 a_2}] = R^{a_1 a_2}  \ , \ [R,R^{a_1 ... a_{22}}] = - R^{a_1 ... a_{22}}  \ , $$
$$
[R,R^{a_1 ... a_{24}}] = 0  \ , \ [R,R^{a_1..a_{23},b}] = 0 \ , \ [R,R^{a_1 .. a_{25},b_1 .. b_{19}}] = - 2 R^{a_1 .. a_{25},b_1 .. b_{19}} \ , \eqno(B.1)$$
$$
[R^{a_1 a_2},R^{b_1 b_2}] = 0 \ , \ [R^{a_1 a_2},R^{b_1 .. b_{22}}] = R^{a_1 a_2 b_1 .. b_{22}} + R^{b_1 ... b_{22}[a_1 , a_2]} \ , \  $$
$$
[R^{a_1 .. a_{22}},R^{b_1 .. b_{22}}] = R^{a_1 .. a_{22} [b_1 b_2 b_3,b_4 .. b_{22}]} \ ,   $$
\par 
The algebra among the negative generators is given by
$$
[K^a{}_b,R_{c_1 c_2}] = - 2 \delta^a{}_{[c_1} R_{|b|c_2]} \ , \ [K^a{}_b,R_{c_1 .. c_{22}}] = - 22 \delta^a{}_{[c_1} R_{|b|c_2 .. c_{22}]} \ , $$
$$
[K^a{}_b,R_{c_1 .. c_{24}}] = - 24 \delta^a{}_{[c_1} R_{|b|c_2 .. c_{24}]} \ , $$
$$
[K^a{}_b,R_{c_1 .. c_{23},d}] = - \delta^a{}_{d} R_{d_1 .. d_{23},b}  - 23 \delta^a{}_{[c_1} R_{|a|c_2 .. c_{23},d]} \ , $$
$$
[R,R_{a_1 a_2}] = - R_{a_1 a_2} \ , \ [R,R_{a_1 ... a_{22}}] = + R_{a_1 ... a_{22}} \ , \ 
[R,R_{a_1 ... a_{24}}] = 0 \ , \ [R,R_{a_1..a_{23},b}] = 0 \ , \ $$
$$
[R_{a_1 a_2},R_{b_1 b_2}] = 0 \ , \ [R_{a_1 a_2},R_{b_1 .. b_{22}}] = R_{a_1 a_2 b_1 .. b_{22}} + R_{b_1 ... b_{22}[a_1 , a_2] } \ . 
 \eqno(B.2)$$
The positive and negative generator algebra is
$$
[R^{a_1 a_2},R_{b_1 b_2}] = 4 \delta^{[a_1}{}_{[b_1} K^{a_2]}{}_{b_2]} - {1 \over 6} \delta^{a_1 a_2}_{b_1 b_2} D + {1 \over 3} \delta^{a_1 a_2}_{b_1 b_2} R  \ \ ,$$
$$
[R^{a_1 a_2},R_{b_1 ... b_{22}}] = [R_{a_1 a_2},R^{b_1 ... b_{22}}] = 0 \ \ ,$$ 
$$
[R^{a_1 .. a_{22}},R_{b_1 .. b_{22}}] = (22)^2 \cdot 21! \delta^{[a_1 .. a_{21}}_{[b_1 .. b_{21}} K^{a_{22}]}{}_{b_{22}]} - {11 \over 12} 22! \delta^{a_1 .. a_{22}}_{b_1 .. b_{22}} D - {22! \over 6} \delta^{a_1 .. a_{22}}_{b_1 .. b_{22}} R  \ \ ,$$
$$
[R^{a_1 a_2},R_{b_1 .. b_{23},c}] = - {23 \cdot 11 \over 3} (\delta^{a_1 a_2}_{[b_1 b_2} R_{b_3 .. b_{23}]c} + \delta^{a_1 a_2}_{c[b_1} R_{b_2 .. b_{23}]}) \ \ , $$
$$
[R^{a_1 a_2},R_{b_1 .. b_{24}}] = 4 \cdot 23 \delta^{a_1 a_2}_{[b_1 b_2} R_{b_3 .. b_{24}]} \ \ , $$
$$
[R^{a_1 .. a_{22}},R_{c_1 .. c_{23},d}] = - {23! \cdot 11 \over 6} (\delta^{a_1 a_2 .. a_{22}}_{d \ [c_1 .. c_{21}} R_{c_{22} c_{23}]} + \delta^{a_1 .. a_{22}}_{[c_1 .. c_{22}} R_{c_{23}]d}) \ \ , 
$$
$$
[R^{a_1 .. a_{22}},R_{b_1 .. b_{24}}] = - {24! \over 12} \delta^{a_1 .. a_{22}}_{[b_1 .. b_{22}} R_{b_{23} b_{24}]} \ \ ,\ 
[R_{a_1 .. a_{22}},R^{b_1 .. b_{24}}] = - {24! \over 12} \delta_{a_1 .. a_{22}}^{[b_1 .. b_{22}} R^{b_{23} b_{24}]} \ \ , $$
$$
[R^{a_1 .. a_{24}},R_{b_1 .. b_{23},c}]  = 0  \ \ , \ \ [R^{a_1 .. a_{24}},R_{b_1 .. b_{24}}] = -  {(24)^2 \over 6} \cdot 23!  \delta^{[a_1 .. a_{23}}_{[b_1 .. b_{23}} K^{a_{24}]}{}_{b_{24}]}  + {24! \over 6} \delta^{a_1 .. a_{24}}_{b_1 .. b_{24}} D \ \ , $$
$$
[R^{a_1 .. a_{23},b},R_{c_1 .. c_{23},d}] = - 23! {23 \over 6}  ( \delta^{a_1 .. a_{23}}_{c_1 .. c_{23}} K^b{}_d + \delta^{[a_1 .. a_{22}|b|}_{c_1 .. c_{22} c_{23}} K^{a_{23}]}{}_d + \delta^{a_1 .. a_{22} a_{23}}_{[c_1 .. c_{22}|d|} K^b{}_{c_{23}]} $$
$$
 + 23 \delta^b{}_d \delta^{[a_1 .. a_{22}}_{[c_1 .. c_{22}} K^{a_{23}]}{}_{c_{23}]} - 22 \delta^{[a_1}{}_d \delta^{a_2 .. a_{22}|b|}_{[c_1 .. c_{21} c_{22}} K^{a_{23}]}{}_{c_{23}]})  + 23! {23 \over 6} (\delta^b{}_d \delta^{a_1 .. a_{23}}_{c_1 .. c_{23}} + \delta^{[a_1}{}_d \delta^{a_2 .. a_{23}]b}_{c_1 .. c_{22} c_{23}}) D 
  \eqno(B.3)$$
\par 

\par 
The algebra of $I_c(K_{27})$ is given as
$$
[J_{a_1 a_2},J_{b_1 b_2}] =  4 \eta_{[a_1|[b_1} J_{b_2]|a_2]} \ \ , \ \ [J_{a_1 a_2},S_{b_1 b_2}] = 4 \eta_{[a_1|[b_1} S_{b_2]|a_2]} \ \ , $$
$$
[J_{a_1 a_2},S_{b_1 .. b_{22}}] = 2 \cdot 22 \eta_{[a_1|[b_1} S_{b_2 .. b_{22}]|a_2]} \ \ , \ \  [J_{a_1 a_2},S_{b_1 .. b_{24}}] = 2 \cdot 24 \eta_{[a_1|[b_1} S_{b_2 .. b_{24}]|a_2]} $$
$$
[J_{a_1 a_2},S_{b_1 .. b_{23},c}] = - 2 \cdot 23 \eta_{[a_1|[b_1} S_{b_2 .. b_{23}]|a_2],c} - 2 \eta_{c[a_1|} S_{b_1 .. b_{23},|a_2]} $$
$$
[S_{a_1 a_2},S_{b_1 b_2}] = + 4 \eta_{[a_1|[b_1} J_{b_2]|a_2]} \ \ , \ \ [S_{a_1 a_2},S_{b_1 .. b_{22}}] = S_{a_1 a_2 b_1 .. b_{22}} + S_{b_1 .. b_{22} [a_1,a_2]} $$
$$
[S_{a_1 a_2},S_{b_1 .. b_{24}}] = + 4 \cdot 23 \eta_{a_1 c_1} \eta_{a_2 c_2}  \delta^{c_1 c_2}_{[b_1 b_2} S_{b_3 .. b_{24}]} $$
$$
[S_{a_1 a_2},S_{b_1 .. b_{23},c}] =  {23 \cdot 11 \over 3}  \eta_{e_1 a_1} \eta_{e_2 a_2} (\delta^{e_1 e_2}_{[b_1 b_2} S_{b_3 .. b_{23}]c} + \delta^{e_1 e_2}_{c[b_1} S_{b_2 .. b_{23}]}) $$
$$
[S^{a_1 .. a_{22}},S_{b_1 .. b_{22}}] = - (22)^2 21! \delta^{[a_1 .. a_{21}}_{[b_1 .. b_{21}} J^{a_{22}]}{}_{b_{22]}} $$
$$
[S_{a_1 .. a_{22}},S_{b_1 .. b_{24}}] =  {24 \over 12!} \eta_{c_1 a_1} .. \eta_{c_{22} a_{22}} \delta^{c_1 .. c_{22}}_{[b_1 .. b_{22}} S_{b_{23} b_{24}]}  $$
$$
[S_{a_1 .. a_{22}},S_{b_1 .. b_{23},c}] =  + {23! \cdot 11 \over 6} (\delta_{c[b_1 .. b_{21}}^{a_1 a_2 .. a_{23}} S_{b_{22} b_{23}]} + \delta_{[b_1 .. b_{22}}^{a_1 .. a_{22}} S_{b_{23}]c}) $$
$$
[S^{a_1 .. a_{24}},S_{b_1 .. b_{24}}] = - {(24)^2 23! \over 6} \delta^{[a_1 .. a_{23}}_{[b_1 .. b_{23}} J^{a_{24}]}{}_{b_{24}]} \ \ , \ \ $$
$$
[S^{a_1 .. a_{23},b},S_{c_1 .. c_{23},d}] = - 23! {23 \over 6} ( \delta^{a_1 .. a_{23}}_{c_1 .. c_{23}} J^b{}_d + \delta^{[a_1 .. a_{22}|b|}_{c_1 .. c_{22} c_{23}} J^{a_{23}]}{}_d + \delta^{a_1 .. a_{22} a_{23}}_{[c_1 .. c_{22}|d|} J^b{}_{c_{23}]} 
$$
$$
+ 23 \delta^b{}_d \delta^{[a_1 .. a_{22}}_{[c_1 .. c_{22}} J^{a_{23}]}{}_{c_{23}]} - 22 \delta^{[a_1}{}_d \delta^{a_2 .. a_{22}|b|}_{[c_1 .. c_{21} c_{22}} J^{a_{23}]}{}_{c_{23}]}) 
\eqno(B.4)$$
\par 
To give an indication of the difficulties encountered when $R^{25,19}$ is included, we note that the commutators involving generators of levels 1 with -1 will have a  commutators involving $R_{b_1 .. b_{25},c_1 .. c_{19}}$ such as 
$$
[R^{a_1 .. a_{22}},R_{b_1 .. b_{25},c_1 .. c_{19}}] =  (A_0 R_{[c_1 .. c_6 | d_1 .. d_{19}| } \delta^{a_1 .. a_{22}}_{c_7 .. c_{25}]} + A_1 R_{[c_1 .. c_7 | [d_1 .. d_{18} } \delta^{a_1 \ a_2 \ .. a_{22}}_{d_{19}]|c_8 .. c_{25}]}   
$$
$$
 \ \ \ + A_2 R_{[c_1 .. c_8 | [d_1 .. d_{17}} \delta^{a_1 \ a_2 \ a_3 \ .. a_{22}}_{d_{18} d_{19}]|c_9 .. c_{25}]} + A_3 R_{[c_1 .. c_9 | [d_1 .. d_{16} } \delta^{a_1 \ a_2 \ a_3 \ a_4 \ .. a_{22}}_{d_{17} d_{18} d_{19}]|c_{10} .. c_{25}]} +
$$
$$
 ... + A_{18} R_{[c_1 .. c_{18} | [d_1 } \delta^{a_1 .. a_2 \ a_3 \ .. a_{22}}_{d_2 .. d_{19}]|c_{19} .. c_{25}]} )
  \eqno(B.5) $$
 The 19 coefficients $A_0, A_1,...,A_{18}$ can be determined by taking Jacobi identities.  We will not need these coefficients in this paper. 


\medskip
{\bf References}
\medskip
\item{[1]} C. Campbell and P. West, {\it $N=2$ $D=10$ nonchiral supergravity and its spontaneous compactification.}
Nucl.\ Phys.\ {\bf B243} (1984) 112;  M. Huq and M. Namazie, {\it Kaluza--Klein supergravity in ten dimensions},
Class.\ Q.\ Grav.\ {\bf 2} (1985); F. Giani and M. Pernici, {\it $N=2$ supergravity in ten dimensions},
Phys.\ Rev.\ {\bf D30} (1984) 325.
\item{[2]} J, Schwarz and P. West, {\it ÊSymmetries and Transformation of Chiral $N=2$ $D=10$ Supergravity}, Phys. Lett. {\bf 126B} (1983) 301;  P. Howe and P. West, {\it The Complete $N=2$ $D=10$ Supergravity}, Nucl.\ Phys.\ {\bf B238} (1984) 181;  J. Schwarz,
{\it Covariant Field Equations of Chiral $N=2$ $D=10$ Supergravity}, Nucl.\ Phys.\ {\bf B226} (1983) 269.
\item{[3]}  E. Cremmer, B. Julia and J. Scherk, {\it Supergravity Theory in Eleven-Dimensions}, Phys. Lett. {\bf 76B} (1978) 409.
\item{[4]} P. Townsend, {\it The eleven dimensional supermembrane revisited}, Phys.\ Lett.\ {\bf 350B} (1995) 184,
arXiv:hep-th/9501068. 
\item{[5]} E. Witten, {\it String theory dynamics in various dimensions}, Nucl.\ Phys.\ {\bf B443} (1995) 85, ÊarXiv:hep-th/9503124.
\item{[6]} P. West, {\it $E_{11}$ and M Theory}, Class. Quant. Grav.  {\bf 18}, (2001) 4443, hep-th/ 0104081.
\item{[7]} P. West, {\it $E_{11}$, SL(32) and Central Charges}, Phys. Lett. {\bf B 575} (2003) 333-342,  hep-th/0307098.
\item{[8]} P. West,{\it A brief review of E theory}, Proceedings of Abdus Salam's 90th  Birthday meeting, 25-28 January 2016, NTU, Singapore, Editors L. Brink, M. Duff and K. Phua, World Scientific Publishing and IJMPA, {\bf Vol 31}, No 26 (2016) 1630043, arXiv:1609.06863.
\item{[9]} A. Tumanov and P. West, {\it E11 must be a symmetry of strings and branes },  arXiv:1512.01644.
\item{[10]} A. Tumanov and P. West, {\it E11 in 11D}, Phys.Lett. B758 (2016) 278, arXiv:1601.03974.
\item{[11]} Kac-Moody Symmetries of IIB Supergravity, (with I. Schnakenburg), 
Phys.Lett. B517 (2001) 421-428, hep-th/0107181.
\item{[12]} P. West, {\it Spacetime and large local transformations },     Int.J.Mod.Phys. {\bf A 38} (2023) 08, 2350045
, arXiv:2302.02199. 
\item{[13]} P. West,  {\it  Irreducible representations of E theory},  Int.J.Mod.Phys. A34 (2019) no.24, 1950133,  arXiv:1905.07324.
\item{[14]} C. G. Callan and L. Thorlacius, {\it Sigma Models And String Theory,} In ÒProvidence 1988, Proceedings, Particles, strings and supernovaeÓ, vol. 2, 795-878; C. Lovelace, {\it Stability Of String Vacua. 1. A New Picture Of The Renormalization
Group,} Nucl. Phys. B 273, 413 (1986). ÒStrings In Curved Space,Ó Phys. Lett. B 135, 75 (1984).
\item{[15]} E. S. Fradkin and A. A. Tseytlin, {\it Effective Field Theory From Quantized Strings,} Phys. Lett. B 158, 316 (1985). ÒQuantum String Theory Effective Action,Ó Nucl. Phys. B 261, 1 (1985); C. G. Callan, E. J. Martinec, M. J. Perry and D. Friedan, {\it Strings In Background Fields,} Nucl. Phys. B 262, 593 (1985). C. G. Callan, I. R. Klebanov and M. J. Perry, ÒString Theory Effective Actions,Ó Nucl. Phys. B 278, 78 (1986).
\item{[16]} P. Cook, {\it Connections between Kac-Moody algebras and M-theory}, ArXiv:0711.3498
\item{[17]} P. West, {\it Introduction to Strings and Branes}, Cambridge University Press, 2012.
\item{[18]} T.  Nutma, {\it SimpLie, a simple program for Lie algebras},  \break 
https://code.google.com/p/simplie/.
\item{[19]}  P. West, {\it Dual gravity and E11},  arXiv:1411.0920. 
\item{[20]} P. West,  {\it On the different formulations of the E11 equations of motion}, Mod.Phys.Lett. A32 (2017) no.18, 1750096,  arXiv:1704.00580. 
\item{[21]} A. Kleinschmidt and P. West, {\it Representations of G+++ and the role of space-time},
JHEP 0402 (2004) 033, hep-th/0312247.; 
P. West, {\it E11 origin of brane charges and U-duality multiplets}, JHEP 0408 (2004) 052, hep-th/0406150.
\item{[22]} Arkady Tseytlin. {\it Duality Symmetric Formulation of String World Sheet Dynamics}. Phys. Lett. B, 242:163Ð174, 1990; 
 {\it Duality symmetric closed string theory and interacting chiral scalars}. Nucl. Phys. B, 350:395Ð440, 1991.
\item{[23]} M. Duff. {\it Duality Rotations in String},  Theory. Nucl. Phys. B, 335:610, 1990.
\item{[24]} P. West, {\it Generalised space-time and duality},   Phys.Lett.B 693 (2010) 373, arXiv:1006.0893 
\item{[25]} Warren Siegel. {\it Superspace duality in low-energy superstrings}. Phys. Rev. D,48:2826Ð2837, 1993;
  Two vierbein formalism for string inspired axionic gravity. Phys.Rev. D, 47:5453Ð5459, 1993. 
\item{[26]}  P. West, {\it E11, generalised space-time and IIA string theory}, Phys.Lett.B 696 (2011) 403-409,   arXiv:1009.2624. 
\item{[27]} A.  Rocen and P. West, {\it E11, generalised space-time and IIA string theory the R-R sector},  in Strings, Gauge fields and the Geometry behind:The Legacy of Maximilian Kreuzer, edited by  Anton Rebhan, Ludmil Katzarkov,  Johanna Knapp, Radoslav Rashkov, Emanuel Scheid, World Scientific, 2013, arXiv:1012.2744.

\end